\documentclass[11pt, a4paper]{article}
\pdfoutput=1

\usepackage[height=8.85in,width=6.75in]{geometry}

\usepackage{graphicx,rotating}     
\usepackage[bookmarksopen,colorlinks=true,linkcolor=light_blue,
citecolor=darkred,urlcolor=darkred,linktoc=all]{hyperref}

\usepackage{amsmath,amssymb,tensor,subdepth}
\usepackage{mathtools}
\usepackage{slashed}
\usepackage{bm}
\usepackage{bbm}
\usepackage{cite}
\usepackage{comment}
\usepackage[utf8]{inputenc}
\usepackage{accents}
\usepackage[footnotesize]{caption}
\usepackage{cancel}
\usepackage{lettrine}
\usepackage{physics}

\usepackage[lf]{Baskervaldx} 
\usepackage[bigdelims,vvarbb]{newtxmath} 
\usepackage[cal=boondoxo]{mathalfa} 

\usepackage{stmaryrd}
\usepackage{trimclip}
\usepackage[framemethod=default]{mdframed}
\newmdenv[skipabove=7pt,
skipbelow=7pt,
rightline=false,
leftline=false,
topline=false,
bottomline=false,
backgroundcolor=gray!10,
linecolor=gray,
innerleftmargin=5pt,
innerrightmargin=5pt,
innertopmargin=5pt,
innerbottommargin=5pt,
leftmargin=0cm,
rightmargin=0cm,
linewidth=4pt]{eBox}
\newmdenv[skipabove=7pt,
skipbelow=7pt,
rightline=true,
leftline=true,
topline=true,
bottomline=true,
backgroundcolor=white,
linecolor=gray,
innerleftmargin=5pt,
innerrightmargin=5pt,
innertopmargin=5pt,
innerbottommargin=5pt,
leftmargin=0cm,
rightmargin=0cm,
linewidth=1pt]{eBox2}

\input Zallman.fd

\LettrineTextFont{\itshape}
\usepackage{chngcntr}
\counterwithin{equation}{section}

\newcommand{\Nd}{N_1}
\newcommand{\Nl}{N_2}
\newcommand{\Nh}{N_3}
\newcommand{\Md}{M_1}
\newcommand{\Ml}{M_2}
\newcommand{\Mh}{M_3}
\newcommand{\xd}{x_1}
\newcommand{\yd}{y_{1 \alpha}}

\usepackage{xcolor}
\definecolor{darkred}{rgb}{0.7, 0., 0.}
\definecolor{orangered}{rgb}{1,0.27,0.}
\definecolor{steelblue}{rgb}{0.275,0.51, 0.706}
\definecolor{forestgreen}{rgb}{0.13,0.55,0.13}
\definecolor{brightgreen}{cmyk}{0.75, 0.02, 1.00, 0.00}
\definecolor{dark_red}{rgb}{0.7, 0., 0.}
\definecolor{light_pink}{rgb}{1,0.4,0.4}
\definecolor{light_blue}{rgb}{0.284602,0.317763,0.963947}
\newcommand{\Mpl}{M_{\text{Pl}}}

\usepackage{tikz}
\usepackage{tikz-feynman}
\tikzfeynmanset{compat=1.0.0}
\pgfdeclarelayer{bg}    
\pgfsetlayers{bg,main}  
\usetikzlibrary{decorations.shapes}
\allowdisplaybreaks

\pgfdeclaredecoration{dashsoliddouble}{initial}{
  \state{initial}[width=\pgfdecoratedinputsegmentlength]{
    \pgfmathsetlengthmacro\lw{.3pt+.5\pgflinewidth}
    \begin{pgfscope}
      \pgfpathmoveto{\pgfpoint{0pt}{\lw}}%
      \pgfpathlineto{\pgfpoint{\pgfdecoratedinputsegmentlength}{\lw}}%
      \pgfmathtruncatemacro\dashnum{%
        round((\pgfdecoratedinputsegmentlength-3pt)/6pt)
      }
      \pgfmathsetmacro\dashscale{%
        \pgfdecoratedinputsegmentlength/(\dashnum*6pt + 3pt)
      }
      \pgfmathsetlengthmacro\dashunit{3pt*\dashscale}
      \pgfsetdash{{\dashunit}{\dashunit}}{0pt}
      \pgfusepath{stroke}
      \pgfsetdash{}{0pt}
      \pgfpathmoveto{\pgfpoint{0pt}{-\lw}}%
      \pgfpathlineto{\pgfpoint{\pgfdecoratedinputsegmentlength}{-\lw}}%
      \pgfusepath{stroke}
    \end{pgfscope}
  }
}

\begin{document}

\hypersetup{pageanchor=false}
\begin{titlepage}

\begin{center}

\hfill KEK-TH-2696 \\ 
\hfill IPMU25-0014 \\
\hfill RESCEU-5/25

\vskip 1in

{\Huge \bfseries
No-scale Brans--Dicke Gravity\\
--ultralight scalar boson \& heavy inflaton--\\
}
\vskip .8in

{\Large Muzi Hong$^{a,b}$, Kyohei Mukaida$^{c}$, Tsutomu T. Yanagida$^{b,d}$} \\

\vskip .3in
\begin{tabular}{ll}
$^{a}$& \!\!\!\!\!\emph{Department of Physics, Graduate School of Science, 
The University of Tokyo, Tokyo 113-0033, Japan}\\
$^{a}$&\!\!\!\!\!\emph{
RESCEU, Graduate School of Science, 
The University of Tokyo, Tokyo 113-0033, Japan}\\
$^{b}$&\!\!\!\!\!\emph{Kavli IPMU (WPI), UTIAS, The University of Tokyo, Kashiwa, Chiba 277-8583, Japan}\\
$^{c}$ & \!\!\!\!\!\emph{Theory Center, IPNS, KEK, 1-1 Oho, Tsukuba, Ibaraki 305-0801, Japan}\\
$^{c}$ & \!\!\!\!\!\emph{Graduate University for Advanced Studies (Sokendai), 
1-1 Oho, Tsukuba, Ibaraki 305-0801, Japan}\\
$^{d}$& \!\!\!\!\!\emph{Tsung-Dao Lee Institute \& School of Physics and Astronomy, Shanghai Jiao Tong University,}\\[-.3em]
& \!\!\!\!\!\emph{Pudong New Area, Shanghai 201210, China}\\
\end{tabular}

\end{center}
\vskip .6in

\begin{abstract}
\noindent

\end{abstract}
It is very much intriguing if the Planck scale $\Mpl$ is not a fundamental parameter. The Brans--Dicke gravity is nothing but the theory where the Planck scale $\Mpl$ is indeed an illusional parameter. The theory predicts a massless scalar boson whose exchanges between matters induce unwanted long range forces. We solve this problem imposing there is no dimensionful parameter in the theory, even at the quantum level. 
We further extend the theory by including a $R^2$ term and a non-minimal coupling of the Standard Model Higgs to gravity, as their coefficients are dimensionless. This extension provides a heavy inflaton field that is consistent with all cosmological observations, with a potential very similar to that of the Starobinsky model. The inflaton necessarily decays into the massless scalar bosons, resulting in a non-negligible amount of dark radiation in the present universe. We demonstrate that the inflation model yields a sufficiently high reheating temperature for successful leptogenesis, and we also discuss a possible candidate for dark matter.

\end{titlepage}

\tableofcontents
\renewcommand{\thepage}{\arabic{page}}
\renewcommand{\thefootnote}{$\natural$\arabic{footnote}}
\setcounter{footnote}{0}
\hypersetup{pageanchor=true}

\section{Introduction}
\label{sec:introduction}
We have three dimensionful parameters in nature, that is, the light velocity $\mathscr{c}$, the Planck constant $\hbar$ and the Planck mass $\Mpl$. The first two constants are fundamental constants related to a space-time symmetry and a principle between the space-time and their conjugate momentum. However, the Planck mass $\Mpl$ is required only to define the strength of the gravitational interactions. Thus, it might be not fundamental, but even an illusional effective parameter.

The Brans--Dicke (BD) gravity~\cite{Brans:1961sx} is nothing but the theory where the Planck scale $\Mpl$ is not fundamental, but indeed an illusional parameter. This theory predicts a massless scalar boson in addition to the graviton. Consequently, the BD theory is already subject to a very strong constraint~\cite{Will:2005va}, since the exchange of the massless or ultralight scalar boson induces too strong long range forces between matters.

In this paper, we solve, first,  the above problem by imposing that the theory does not have any dimensionful parameters beside the light velocity $\mathcal{c}$ and the Planck constant $\hbar$ . We write down explicitly the standard model (SM) Lagrangian without introducing dimensionful parameters including the Majorana mass term for the right-handed neutrinos $N$. We call it \textit{no-scale BD gravity}. And we show that the scalar boson is completely decoupled from the SM particles as expected~\cite{Ferreira:2016kxi, Burrage:2018dvt}, even at the quantum theory level, and hence its exchange never generates the dangerous long range force.

Furthermore, we extend the no-scale BD gravity to include the $R^2$ term and the $\abs{H}^2 R$ term with the SM Higgs being $H$, since their coefficients are dimensionless.\footnote{
    We forbid the $R_{\mu\nu}R^{\mu\nu}$ term, since it makes the theory non-unitary.
}
Then, we find an additional scalar boson which can be identified with an inflaton~\cite{Rinaldi:2015uvu, Tambalo:2016eqr,Ferreira:2018qss}.
We show, in this paper, that the inflationary universe based on this scenario is perfectly consistent with all present cosmological observations; such as the inflationary observables, the neutrino masses, the baryon asymmetry of the universe, and the amount of dark matter (DM).
The inflaton necessarily decays to the massless scalar bosons discussed above, and hence we have non-negligible dark radiation (DR) in the present universe.
It could be a smoking gun for the present model (see the discussion in the final section of this paper).
As for the DM, we identify one of the right-handed neutrinos with the DM~\cite{Kusenko:2010ik,Cox:2017rgn}.
In the final section, we provide our conclusions in this paper and discussion on future possible problems. We take the natural unit, $\mathscr{c}=\hbar=1 $ throughout this paper.

\section{No-scale Brans--Dicke gravity}
\label{sec:no-scale-BD}
The original BD gravity~\cite{Brans:1961sx} assumes its Lagrangian as
\begin{equation}
    \mathcal{L} =
    \frac{\xi}{2}\phi^2 R + \frac{1}{2} \qty(\partial \phi)^2 - \frac{\lambda}{4} \phi^4
    + \mathcal{L}_\text{SM},
\end{equation}
where we denote the standard model (SM) Lagrangian as $\mathcal{L}_\text{SM}$ and assume $\xi > 0$ to obtain the correct sign of the gravitational coupling.

One may always perform the following Weyl transformation to go to the Einstein frame
\begin{equation}
    \label{eq:weyl}
    g_{\mu\nu}^\text{E} = \Omega^2 g_{\mu\nu}, \qquad \Omega^2 = \frac{\xi \phi^2}{\Mpl^2},
\end{equation}
which induces the transformation of the Ricci scalar as
\begin{equation}
    \label{eq:weyl_Ricci}
    R = \Omega^2 \qty( R_\text{E} - 3 \Box_\text{E} \ln \Omega^2 + \frac{3}{2} g^{\mu\nu}_\text{E} \partial_\mu \ln \Omega^2 \partial_\nu \ln \Omega^2).
\end{equation}
Plugging the transformation \eqref{eq:weyl} and \eqref{eq:weyl_Ricci} into the original Lagrangian, we obtain the Einstein frame Lagrangian as
\begin{equation}
    \mathcal{L}_\text{E} = \frac{\Mpl^2}{2} R_\text{E} + \frac{1}{2} \qty(\partial \chi)^2 - \frac{\lambda}{4 \xi^2} \Mpl^4 + \Omega^{-4}\mathcal{L}_\text{SM},
\end{equation}
where the canonically normalized massless scalar field $\chi$ is defined as
\begin{equation}
    \chi = \sqrt{1 + \frac{1}{6 \xi} } \, \sqrt{\frac{3}{2}} \Mpl  \ln \frac{\xi \phi^2}{\Mpl^2} ~\longrightarrow~\Omega^2 = e^{\sqrt{\frac{2}{3}} \sqrt{\frac{6 \xi}{6 \xi + 1} } \frac{\chi}{\Mpl}},
\end{equation}
and the fields in $\mathcal{L}_\text{SM}$ are redefined such that they have canonical kinetic terms.
The smallness of the cosmological constant is now translated to the smallness of the dimensionless parameter $\lambda/ \xi^2$ at the current scale.
Note that the singular point at $\phi=0$ in the Weyl transformation [Eq.~\eqref{eq:weyl}] is now mapped to the limit $\chi\to-\infty$.
Hence, one may practically forget about this as long as we consider a finite field value of $\chi$.

We take a closer look at the interaction between the scalar field $\chi$ and the SM particles.
The tree level interactions are induced if the SM Lagrangian involves non-conformally coupled fields, such as the Higgs boson.
The Higgs mass term, $m_H^2 |H|^2$, in the SM Lagrangian yields the following interaction term in the Einstein frame:
\begin{equation}
    \Omega^{-4}\mathcal{L}_\text{SM} \supset m_H^2 \abs{H}^2 \Omega^{-2} \to m_H^2 \abs{H}^2 e^{-\sqrt{\frac{2}{3}} \sqrt{\frac{6 \xi}{6 \xi + 1} } \frac{\chi}{\Mpl}},
\end{equation}
where we have redefined $H \to \Omega H$ so that the Higgs field is canonically normalized.\footnote{
    This also induces the kinetic mixing between the Higgs and $\chi$ unless the Higgs has the conformal coupling to the Ricci scalar. However, the kinetic mixing does not introduce any dangerous long range force as we will see later in this section. 
}
This operator induces the mixing between the Higgs boson and $\chi$ after the electroweak symmetry breaking as $ (v / \Mpl) m_H^2 h \chi $ with $h$ being the Higgs mode after the electroweak symmetry breaking, and hence $\chi$ couples to all the SM particles, which is excluded by the stringent constraint on the fifth force experiments (see \textit{e.g.,} \cite{Adelberger:2003zx}).

Now we propose an extension of the original BD gravity so that there is no fundamental dimensionful parameter.
In other words, all the dimensionful parameters are measured with respect to $\sqrt{\xi}\phi$, above which the gravity becomes strongly coupled in the original Lagrangian.
This requirement not only implies the trivial replacement of the dimensionful parameters by $\phi$ but the scale associated with the renormalization should also be determined with respect to $\phi$.
Let us first discuss the replacement of the dimensionful parameters.
In the SM Lagrangian, the only dimensionful parameter is the Higgs mass term, which is now expressed as
\begin{equation}
    \mathcal{L}_\text{SM} \supset \lambda_{m_H} \phi^2 \abs{H}^2.
\end{equation}
If we have the Majorana right-handed  neutrino mass term, it is also replaced as
\begin{equation}
    \frac{1}{2} \lambda_{M} \phi N N + \text{H.c.},
\end{equation}
with $N$ being the right-handed Majorana neutrino.
It is straightforward to confirm that these operators never introduce the direct coupling between $\chi$ and the SM particles in the Einstein frame:
\begin{equation}
    \Omega^{-4}\mathcal{L}_\text{SM} \supset \frac{\lambda_{m_H}}{\xi} \Mpl^2 \abs{H}^2 - \frac{1}{2} \qty(\frac{\lambda_{M}}{\sqrt{\xi}} \Mpl N N + \text{H.c.}),
\end{equation}
where we have canonically normalized the right-handed Majorana neutrino in the Einstein frame by $N \to \Omega^{3/2} N$.\footnote{
    Note that the kinetic term of the Weyl fermions and the gauge fields are conformally invariant.
}
The dangerous tree level mass mixing between the Higgs and $\chi$ is now absent as expected.

As advocated earlier, however, this is not enough to suppress the long range force mediated by $\chi$.
Since we are dealing with quantum field theory, any dimensionless couplings can run, implying the dimensionful scale appearing owing to the quantum corrections.
This is not just a technical issue but a practical one, because $\chi$ can couple to all the SM particles through this effect, again leading to the potential pitfalls of the fifth force constraints.
For instance, if we specify the Yukawa coupling  in the original Lagrangian at a certain scale $\Lambda$ as $y (\Lambda)$ in the Jordan frame, the Yukawa coupling in the Einstein frame involves $\chi$ as $y(\Lambda\, e^{-A \chi/\Mpl })$ with $A = \sqrt{1/6} \sqrt{6\xi/(6 \xi + 1)}$.
The same logic applies to all the SM couplings, such as the gauge couplings~\cite{Wetterich:1987fm} and the Higgs quartic coupling, leading to the coupling between $\chi$ and the SM particles via the beta function~\cite{Burrage:2018dvt}.

To avoid this issue, we propose to determine the renormalization scale with respect to $\phi$, known as the scale invariant prescription in the literature~\cite{Englert:1976ep, Shaposhnikov:2008xi,Armillis:2013wya,Hamada:2016onh,Falls:2018olk}.
This may sound artificial, but one can interpret this procedure in the following way.
In the original Lagrangian, the gravity becomes strongly coupled above the scale $\sqrt{\xi}\phi$.
Since we do not intend to deal with quantum gravity, we should not probe the physics larger than the scale $\sqrt{\xi}\phi$.
This may imply that the cutoff scale of the theory should be determined with respect to $\phi$ as $\Lambda /(\sqrt{\xi} \phi) = c$ with $c$ being a constant slightly smaller than unity.
For a scale smaller than the cutoff, $\mu < \Lambda$, one may compute the quantum corrections of couplings in $\mathcal{L}_\text{SM}$ in the usual procedure.
To discuss the low-energy behavior, namely $\mu < \Lambda$, we need to specify the boundary condition of the running couplings at the scale $\Lambda$.
Importantly, this boundary condition of an arbitrary coupling $\lambda_\bullet$ is $\lambda_\bullet (\Lambda)  = \lambda_\bullet (c \sqrt{\xi}\phi)$, depending on $\phi$.
By performing the Weyl transformation to go to the Einstein frame, the boundary condition for all the running couplings is now dictated as $\lambda_\bullet (\Lambda / \Omega) = \lambda_\bullet (c \Mpl)$, leading to the absence of the coupling between the SM fields and $\chi$ via quantum corrections.\footnote{This important point is not fully addressed in the previous papers~\cite{Ferreira:2016kxi, Burrage:2018dvt}.}

One may also rephrase this prescription as follows. The quantization is necessarily associated with a certain dimensionful scale; for instance the cutoff scale in the Pauli--Villars regularization or the renormalization scale in the dimensional regularization.
In the scale invariant prescription, these scales are also replaced by $\phi$, and hence the scale invariance is preserved at the quantum level, where the scale transformation is defined as $\phi \mapsto \phi \omega$, $O \mapsto O \omega^c$ and $g_{\mu\nu} \mapsto g_{\mu\nu} \omega^{-2}$ with $O$ being a field with Weyl weight $c$ and $\omega$ being a constant.

Therefore, the Jordan-frame action defined in this way is equivalent to the Einstein-frame action with all the running couplings being specified at $c \Mpl$, resulting the exact shift-symmetric scalar boson $\chi$ even at the quantum level.

Finally, we comment on the kinetic mixing of the Higgs and $\chi$, which is an inevitable consequence unless the Higgs is conformally coupled to gravity. The kinetic mixing arises from
\begin{equation}
    \Omega^{-4}\mathcal{L}_\text{SM} \supset \frac{1}{\Omega^2} \, g^{\mu\nu}_\text{E} \partial_\mu H^\dag \partial_\nu H
    \to \abs{\partial H}^2
    + \frac{A}{\Mpl} \partial \abs{H}^2 \partial \chi
    + A^2 \frac{\abs{H}^2}{\Mpl^2} \qty( \partial \chi )^2,
\end{equation}
where we have redefined $H \to \Omega H$.
Expanding the Higgs field around the vacuum expectation value (VEV), $|H|^2 \simeq v^2 / 2 + v h + \cdots$, one might wonder whether the mixing between $\chi$ and $h$ induces a dangerous long range force.
As is well known, this never causes such an issue because the coupling induced by the mixing term $ h\Box \chi$ cancels the pole in the $\chi$ propagator, leading to the absence of the long range force.
One may confirm this more explicitly by redefining the $\chi$ field to absorb the kinetic mixing as~\cite{Fukuda:1974kn}
\begin{equation}
    \chi = \bar\chi - \frac{\Mpl}{2 A} \ln \qty( 1 + 2 \frac{A^2 \abs{\bar{H}}^2}{\Mpl^2} ), \qquad H = \bar H,
\end{equation}
where the physical modes $\bar\chi$ and $\bar H$ do not have the kinetic mixing.
Since the original $\chi$ has the shift symmetry, the redefined $\bar\chi$ retains this symmetry, \textit{i.e.,} this redefinition never introduces any problematic couplings to the SM, as expected.
This redefinition removes the kinetic mixing as
\begin{equation}
    \frac{1}{2} \qty(  1 + 2\frac{A^2 \abs{H}^2}{\Mpl^2} ) \qty( \partial \chi )^2
    + \abs{\partial H}^2 - \frac{1}{2} \frac{A^2 \qty( \partial \abs{H}^2 )^2 / \Mpl^2 }{1+2A^2 \abs{H}^2/\Mpl^2},
\end{equation}
where we have dropped the bars on $\chi$ and $H$ for notational brevity.
Expanding the Higgs field around the VEV, one may immediately see that the mixing term between $\chi$ and $h$ is now absent, and hence the long range force is not generated by the kinetic mixing as long as $\chi$ is shift symmetric~\cite{Fukuda:1974kn}.

\section{Inflaton in an extension of the no-scale Brans--Dicke gravity}
\label{sec:inf_noscale}

In general, there is no reason to drop the $R^2$ term and the non-minimal coupling $\xi_H \abs{H}^2 R$ in our construction.
In this section, as a warm-up, we turn off the non-minimal coupling for simplicity, and discuss the effect of the $R^2$ term on inflation, subsequent reheating, dark radiation, and leptogenesis.
The effect of the non-minimal coupling between the Higgs and the Ricci curvature will be discussed in Sec.~\ref{sec:h2R}.

By introducing an auxiliary field $X$, a theory with the $R^2$ term can be written as
\begin{align}
    \mathcal{L} 
    &= \frac{\xi}{2} \phi^2 R + \frac{1}{2} \qty(\partial \phi)^2 - \frac{\lambda}{4} \phi^4 + \alpha R^2 + \mathcal{L}_\text{SM} \\
    &\rightarrow  \qty( \frac{\xi}{2} \phi^2 + 2 \alpha X ) R - \alpha X^2 + \frac{1}{2} \qty(\partial \phi)^2 - \frac{\lambda}{4} \phi^4 + \mathcal{L}_\text{SM}.
\end{align}
One may readily see that the original theory is recovered by integrating out the auxiliary field $X$.
By performing the Weyl transformation of
\begin{equation}
    \Omega^2 = \frac{\xi \phi^2 + 4 \alpha X}{\Mpl^2},
\end{equation}
we may go to the Einstein frame:
\begin{equation}
    \label{eq:einstein_R2}
    \mathcal{L} = \frac{\Mpl^2}{2} R_\text{E} + \frac{3}{4} \Mpl^2 \qty(\partial \ln \Omega)^2 + \frac{1}{2\Omega^2} \qty( \partial \phi )^2
    - \Omega^{-4} \qty( \frac{\lambda}{4} \phi^4 + \alpha X^2 )
    + \Omega^{-4}\mathcal{L}_\text{SM},
\end{equation}

This theory involves two scalar modes where one of them is massless and the other is massive.
To make this property manifest, we introduce the following field redefinition~\cite{Casas:2017wjh}:
\begin{equation}
    \label{eq:field-redef-R2}
    \chi \equiv \sqrt{\frac{3}{2}} \Mpl \ln \frac{\qty( 1 + 6 \xi )\phi^2 + 24 \alpha X}{\Mpl^2}, \qquad \Theta \equiv \frac{\qty(\xi + 1/6) \phi^2 + 4 \alpha X}{\xi \phi^2 + 4 \alpha X}.
\end{equation}
One may rewrite the conformal factor $\Omega^2$ and the scalar field $\phi$ in terms of $\chi$ and $\Theta$ as
\begin{equation}
    \label{eq:Omega-phi-to-chi-Theta}
    \frac{\phi^2}{\Mpl^2} = \qty( 1 - \Theta^{-1} ) e^{ \sqrt{\frac{2}{3}} \frac{\chi}{\Mpl} }, \qquad \Omega^2 = \frac{1}{6\Theta} e^{  \sqrt{\frac{2}{3}} \frac{\chi}{\Mpl} }.
\end{equation}
The singularity of the Weyl transformation at $\Omega^2 \to 0$ is now mapped to $\Theta \to \infty$ or $\chi \to - \infty$.
Hence, again, we do not have to care this singularity as long as we consider a finite field value of $\chi$.
Note also that the former limit is practically forbidden because of the diverging potential energy of $\Theta$ as we will see shortly.

Plugging Eq.~\eqref{eq:Omega-phi-to-chi-Theta} into the Lagrangian \eqref{eq:einstein_R2}, we obtain
\begin{align}
    \label{eq:R2_kin}
    \mathcal{L} &=
    \frac{\Mpl^2}{2} R_\text{E}
    + \frac{1}{2} \Theta \qty(\partial \chi)^2
    + \frac{3}{4} \Mpl^2 \, \frac{1}{\Theta \qty( \Theta - 1 )} \qty(\partial \Theta)^2 \\
    &\quad - 9 \Mpl^4 \qty( \lambda + \frac{\xi^2}{4 \alpha} ) \qty( \Theta - \Theta_0 )^2 
    - \frac{1}{4} \frac{\lambda}{\xi^2 + 4 \alpha \lambda} \Mpl^4
    + 36 \Theta^2 e^{ \sqrt{\frac{2}{3}} \frac{2\chi}{\Mpl}} \mathcal{L}_\text{SM},
    \label{eq:R2_pot}
\end{align}
where
\begin{equation}
    \label{eq:R2_Theta0}
    \Theta_0
    \equiv 1 + \frac{1}{6} \frac{\xi}{4\alpha \lambda + \xi^2}
    \simeq 1 + \frac{1}{6 \xi}.
\end{equation}
Note that the allowed field range of $\Theta$ is $\Theta \geqslant 1$ as can be seen from Eq.~\eqref{eq:field-redef-R2}.
The second term of Eq.~\eqref{eq:R2_pot} is responsible for the current dark energy, whose coefficient must be extremely small.\footnote{
    \label{fn:cc_running}
    Strictly speaking, since there is a running of this coupling, its value is different from the current one at the inflation scale. Still, the value of $\lambda \Mpl^4 / \xi^2$ is extremely smaller than the inflation scale and hence we may neglect this term in the inflationary epoch.
    Note that, to have the correct value of the cosmological constant at the current universe, the boundary condition of the renormalization group flow of this coupling at the inflation scale should be extremely fine-tuned, which is nothing but the fine-tuning problem of the cosmological constant in this context.
}
The current value of $\Theta$ is $\Theta_0$ as can be seen from Eq.~\eqref{eq:R2_pot}, which can be approximated as the last equation of Eq.~\eqref{eq:R2_Theta0} owing to the suppressed value of $\lambda / (4 \xi^2) $.

As emphasized in the previous Sec.~\ref{sec:no-scale-BD}, our central assumption throughout this paper is that all the dimensionful scales must be measured with respect to $\phi$, \textit{i.e.,} not only just dimensionful parameters but the renormalization scale, or equivalently, the cutoff scale of the theory.
After the Weyl transformation, the $\phi$ field acquires the conformal factor as $\phi / \Omega$.
In the model of the previous section, this term $\phi / \Omega$ is just a constant proportional to $\Mpl$ [see Eq.~\eqref{eq:weyl}], and yields the dimensionful parameters in the SM after the Weyl transformation.
Contrary to the model in the previous section, $\phi/\Omega$ in this case is not just a constant, rather it depends on fields as follows:
\begin{equation}
    \frac{\phi^2}{\Omega^2} = 6 \Mpl^2 \qty( \Theta - 1 ).
\end{equation}
One may confirm that this coupling does not involve $\chi$, and hence the $\chi$ field has the shift symmetry.
While the massless scalar $\chi$ is completely decoupled from the SM particles, the massive scalar boson $\Theta$ is coupled to the SM particles through this term.
This fact plays a crucial role when we come to discuss the reheating after inflation.

\subsection{Inflation and reheating}
\label{sec:inf_ref_R2}

\begin{figure}[h]
  \centering
  \includegraphics[width=0.43\columnwidth]{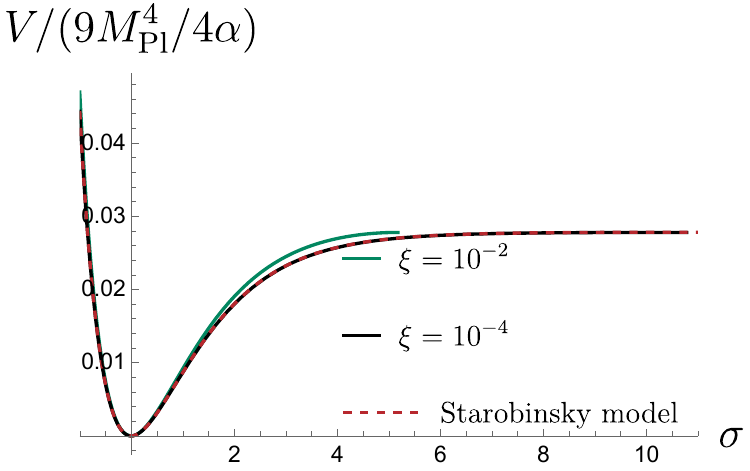}
  \caption{Comparison between the potential in (\ref{eq:inflation_V1}) (black line for $\xi=10^{-4}$ and green line for $\xi=10^{-2}$) and that in (\ref{eq:inflation_V2}) with $\xi=10^{-4}$ (red dashed line), with $\Mpl=1$ and $\xi^2 \gg \alpha \lambda$.}
  \label{fig:potential}
\end{figure}

\begin{figure}[h]
  \centering
  \begin{minipage}{0.43\columnwidth}
    \centering
    \includegraphics[width=\columnwidth]{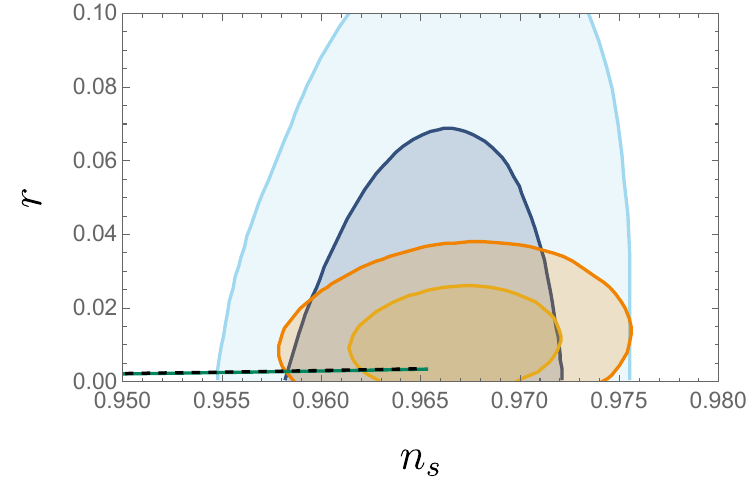}
  \end{minipage}
  \hspace{5mm}
  \begin{minipage}{0.43\columnwidth}
    \centering
    \includegraphics[width=\columnwidth]{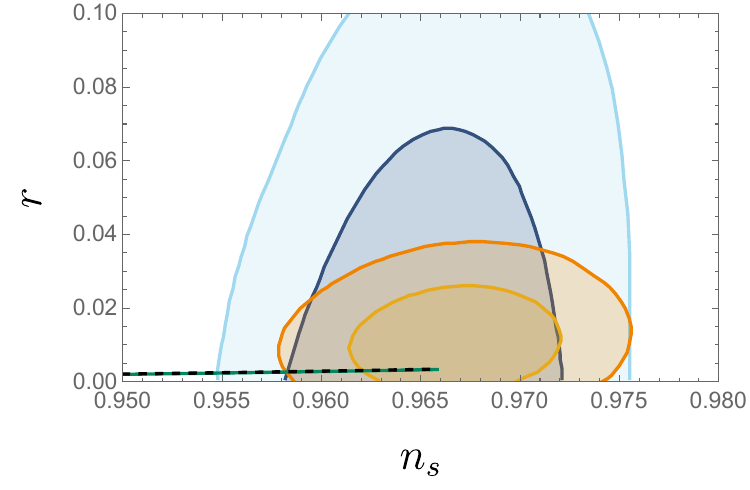}
  \end{minipage}

  \vspace{3mm}
  
  \begin{minipage}{0.43\columnwidth}
    \centering
    \includegraphics[width=\columnwidth]{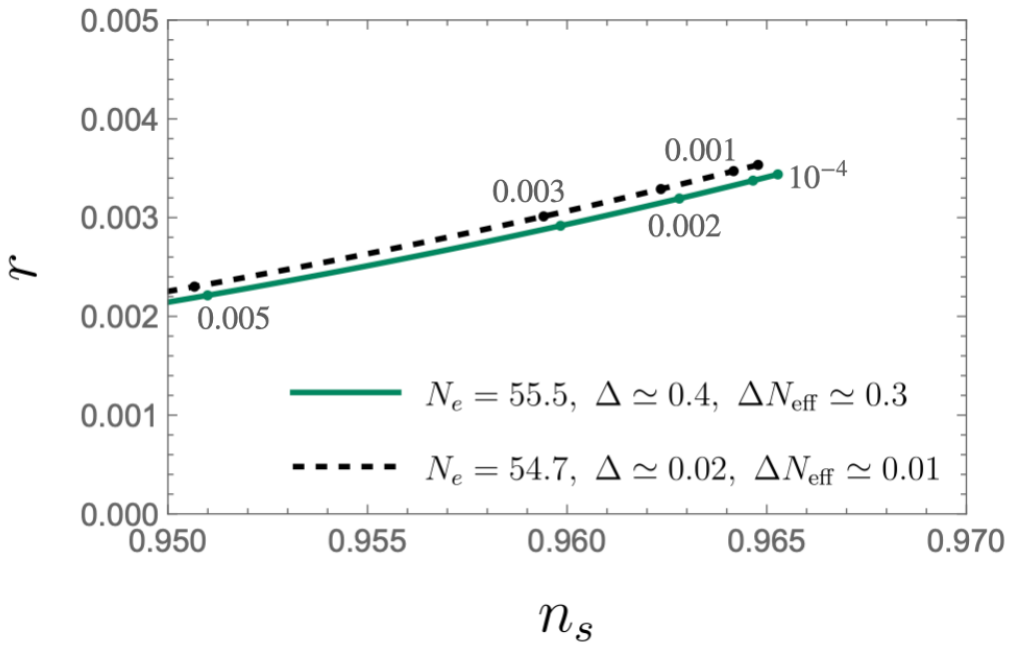}
  \end{minipage}
  \hspace{5mm}
  \begin{minipage}{0.43\columnwidth}
    \centering
    \includegraphics[width=\columnwidth]{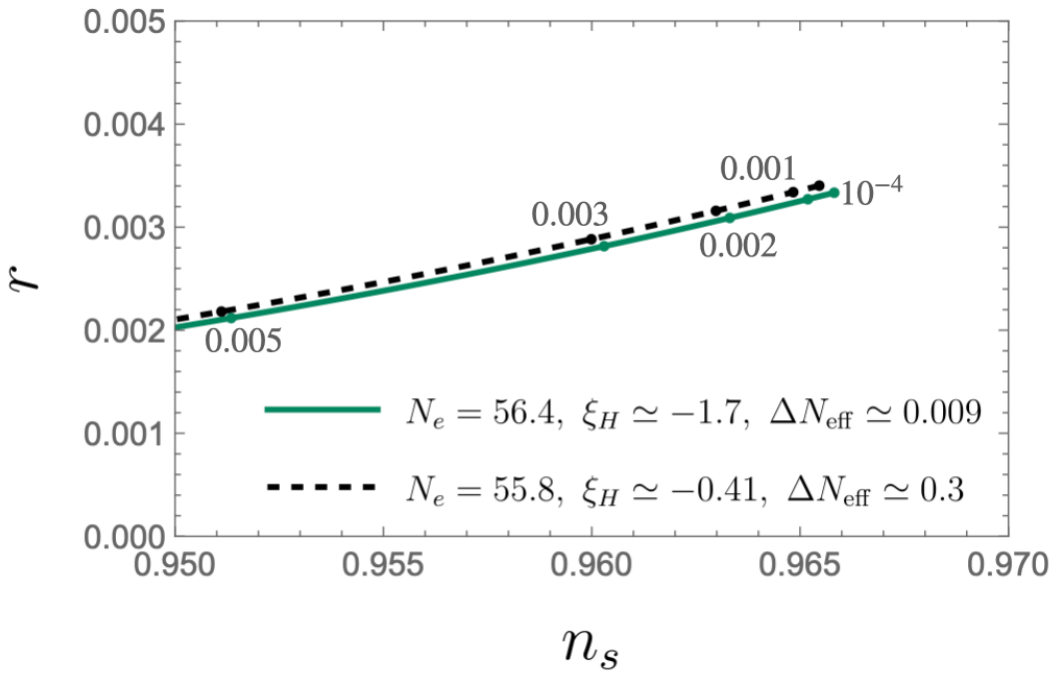}
  \end{minipage}
  \caption{Predictions of tensor-to-scalar ratio $r$ and spectral index $n_s$ of the inflation model (\ref{eq:inflation_model}) with $\Theta_0 \simeq 1 + 1/(6 \xi)$.
  Along the green line and the dashed black line, $\xi$ is taken from $1 \times10^{-5}$, which is the Starobinsky model limit, to $0.05$, which is out of the figures above.
  The blue contours are constraints from Planck \cite{Planck:2018vyg}
  \label{fig:nsr}, and the yellow contours are that from BICEP/Keck \cite{BICEP:2021xfz}.
  (Left) E-fold numbers are taken as borderlines of possible value for the scenario in Sec.~\ref{sec:inf_noscale}. The reheating temperature is fixed as $T_{\rm{R}}=5\times10^9\,{\rm{GeV}}$, and $\Delta$ is taken as $0.02$ and $0.4$ for the corresponding e-fold number $54.7$ and $55.5$, respectively.
  The values of $\Delta$ between $0.02$ and $0.4$ satisfy the restrictions from baryon-to-entropy ratio and that from $\Delta N_{\rm{eff}}$.
  The bottom panel is a zoomed up version of the predictions in the top panel, and a set of points corresponding to $\xi=10^{-4},\ 0.001,\ 0.002,\ 0.003,\ 0.005$ are explicitly shown, which is the same for the right panels.
  (Right) E-fold numbers are taken as borderlines of possible value for the scenario in Sec.~\ref{sec:h2R}.
  $\Delta=1$, and $\xi_H$ is taken as $-1.7$ and $-0.41$  for the corresponding e-fold number $56.4$ and $55.8$, respectively.
  The corresponding temperature is calculated using (\ref{eq:TR_sec4}).
  The values of $\xi_H$ between $-1.7$ and $-0.41$ satisfy the restrictions from the stability of the electroweak vacuum and that from $\Delta N_{\rm{eff}}$.
  }

\end{figure}

Owing to the pole at $\Theta  = 1$, the kinetic term is enhanced in the limit of $\Theta \to 1 +$ and the potential becomes flattened around that point, which opens up the possibility of the slow-roll inflation.
The canonically normalized inflaton field, $\sigma$, is given by
\begin{equation}
    \Theta = \cosh^2 \qty( \sqrt{ \frac{2}{3} } \frac{\sigma}{2\Mpl} ),
\end{equation}
whose field range can be taken as either $\sigma > 0$ or $\sigma < 0$.
From now on we take $\sigma < 0$.
The potential of the inflaton field in terms of $\sigma$ is expressed as
\begin{equation}
    V(\sigma) = 9 \Mpl^4 \qty( \lambda + \frac{\xi^2}{4 \alpha} ) \qty[\cosh^2 \qty( \sqrt{ \frac{2}{3} } \frac{\sigma}{2\Mpl} ) - \Theta_0 ]^2.
    \label{eq:inflation_model}
\end{equation}
As long as $\Theta_0$ is sufficiently large, \textit{i.e.,} $\xi \lesssim 10^{-3}$, we have an enough range of a plateau-like inflaton potential which have similar predictions as the Starobinsky inflation.
To explicitly see the property of the shape of the potential, we shift it so that the minimal point of the potential is at the origin:
\begin{equation}
    V(\sigma) = 9 \Mpl^4 \qty( \lambda + \frac{\xi^2}{4 \alpha} ) \qty[\cosh^2 \qty( \sqrt{ \frac{2}{3} } \frac{\sigma + \sigma_{\rm{min}}}{2\Mpl} ) - \Theta_0 ]^2,
    \label{eq:inflation_V1}
\end{equation}
where
\begin{equation}
    \sigma_{\rm{min}} \equiv
    -\sqrt{6} \Mpl \  {\rm{log}}
    \qty(\sqrt{-1 + 2 \Theta_0 + 2 \sqrt{(-1 + \Theta_0) \Theta_0}}).
\end{equation}
Taking $\Theta_0 \gg 1$, for certain range around the origin, the potential can be approximated by
\begin{equation}
    V(\sigma) \simeq 9 \Mpl^4 \qty( \lambda + \frac{\xi^2}{4 \alpha} ) \Theta_0^2 \qty[1 - {\rm{exp} \qty(- \sqrt{\frac{2}{3}} \frac{\sigma}{\Mpl})}]^2,
    \label{eq:inflation_V2}
\end{equation}
which is the potential of Starobinsky inflation \cite{Starobinsky:1980te, Vilenkin:1985md, Maeda:1987xf}.
A comparison between the shape of the potential above with that of the Starobinsky inflation is shown in Fig.~\ref{fig:potential}.
As can be seen from Fig.~\ref{fig:nsr}, the inflation of this model with suitable parameters is consistent with observation.
Notice that we may have a slightly smaller tensor-to-scalar ratio $r$ than the prediction of the Starobinsky inflation model. 

After the inflation, the inflaton oscillates around the minimum of the potential and decays into the SM particles and $\chi$.
The inflaton potential in this regime can be approximated as
\begin{equation}
    V(\sigma) \simeq \frac{1}{2} m_\sigma^2 \sigma^2, \qquad m_\sigma^2 \equiv 12 \qty( \lambda + \frac{\xi^2}{4\alpha} ) \Theta_0^2 \Mpl^2 \simeq \frac{1}{12 \alpha} \Mpl^2.
\end{equation}
Here we have redefined $\sigma$ so that the minimum of the potential becomes $\sigma = 0$ as above.
Note that by matching the scalar power spectrum amplitude at the CMB scale \cite{Planck:2018vyg}, one has $\alpha \simeq 5 \times 10^8$ and $m_\sigma \simeq 3 \times 10^{13}\,{\text{GeV}}$ for $\xi \ll 1$.
As we will see in the following, since all the coupling to the SM and $\chi$ particles are suppressed by the Planck scale, it takes a long time for the inflaton to complete the reheating.
Hence, the preheating becomes inefficient by the cosmic expansion before the completion of reheating, and the inflaton coherent oscillation dominates the Universe as it behaves as matter.
Eventually, the inflaton completely decays into the SM and $\chi$ particles, whose rate is controlled by the perturbative decay of inflaton into the SM and $\chi$ particles.

After the canonical normalization of $\chi$, one may rewrite the inflaton coupling to $\chi$ in Eq.~\eqref{eq:R2_kin} as
\begin{equation}
    \label{eq:int_sigma_chi}
    \mathcal{L}_{\sigma\chi} \simeq  - \sqrt{\frac{2}{3}} \frac{\sigma}{\Mpl} \frac{1}{2} \qty( \partial \chi )^2,
\end{equation}
where we only keep the leading term in $\sigma /\Mpl$, and use $\Theta_0 \gg 1$.
The decay rate of inflaton to $\chi$ is then given by
\begin{equation}
    \Gamma_{\sigma \to \chi \chi} = \frac{1}{192 \pi} \frac{m_\sigma^3}{\Mpl^2}.
\end{equation}
The couplings to the SM fields are dominated by the kinetic mixing between the Higgs and the mass term of the Majorana right-handed  neutrino because the other couplings are only induced by the quantum corrections, \textit{i.e.,} via the beta functions, which is suppressed compared to the tree level couplings to Higgs and Majorana neutrino.
The coupling to the Higgs is originated from the kinetic mixing between the Higgs and $\sigma$:
\begin{equation}
    \label{eq:int_sigma_H}
    \mathcal{L}_{\sigma H} \simeq -\sqrt{\frac{2}{3}} \frac{\sigma}{\Mpl} \abs{\partial H}^2,
\end{equation}
where we have dropped the Higgs mass as it is negligibly small compared to the inflaton mass.
Again we only keep the leading term in $\sigma /\Mpl$ and use $\Theta_0 \gg 1$.
One may readily see that this coupling is essentially the same as Eq.~\eqref{eq:int_sigma_chi} and hence the decay rate of inflaton to Higgs just differs by the number of degrees of freedom:
\begin{equation}
    \Gamma_{\sigma \to H H^\ast} \simeq \frac{1}{48 \pi} \frac{m_\sigma^3}{\Mpl^2}.
\end{equation}
Here we drop the phase-space factor and the additional coupling since the Higgs mass is negligibly small compared to the inflaton mass.
The interaction between the inflaton and the right-handed Majorana neutrino is originated by the mass term of the right-handed Majorana neutrino:
\begin{equation}
    \label{eq:int_sigma_N}
    \mathcal{L}_{\sigma N} \simeq \sqrt{\frac{1}{6}} \frac{\sigma}{\Mpl} \frac{1}{2} \sum_\alpha M_{\alpha} \qty( N_\alpha N_\alpha + \text{H.c.} ),
\end{equation}
at the leading order in $\sigma/\Mpl$ and $\Theta_0 \gg 1$.
The Majorana mass is given by $M_{N_\alpha} = \lambda_{M_\alpha} \xi^{-1/2} \Mpl $.
The decay rate of inflaton to the Majorana right-handed neutrino is then given by (see \textit{e.g.,} Ref.~\cite{Jeong:2023zrv})
\begin{equation}
    \Gamma_{\sigma \to N_\alpha N_\alpha} = \frac{1}{96 \pi} \frac{m_\sigma^3}{\Mpl^2}
    \times \qty(\frac{M_\alpha^2}{m_\sigma^2})
    \qty( 1 - \frac{4 M_\alpha^2}{m_\sigma^2} )^{3/2}.
\end{equation}
This rate becomes maximum at $M_\alpha = m_\sigma / \sqrt{10}$ as
\begin{equation}
    \Gamma_{\sigma \to N_\alpha N_\alpha} \leqslant \frac{1}{5} \qty( \frac{3}{5} )^{3/2} \Gamma_{\sigma \to \chi \chi} \simeq 0.1 \times \Gamma_{\sigma \to \chi \chi}.
\end{equation}
The branching ratio of the inflaton decay to one Majorana right-handed neutrino is at most $2$\% of the total decay rate.
The inflaton can also couple to the SM gauge bosons via the trace anomaly that is suppressed by the loop factor as\footnote{
    The trace is trivial for $\text{U}(1)_Y$.
} \cite{Gorbunov:2012ns}
\begin{equation}
    \label{eq:int_sigma_gauge}
    \mathcal{L}_{\sigma A} \simeq - \sqrt{\frac{2}{3}} \sum_G \frac{b_G \alpha_G}{8 \pi} \frac{\sigma}{\Mpl} \tr\qty[ F_{\mu\nu} F^{\mu\nu} ],
\end{equation}
where the summation is taken over the SM gauge group $G = \text{SU}(3)_C, \text{SU}(2)_L, \text{U}(1)_Y$, and $b_G$ is the corresponding beta function coefficient of the gauge group $G$ with $b_{\text{SU}(3)_C} = -7$, $b_{\text{SU}(2)_L} = -19/6$, and $b_{\text{U}(1)_Y} = 41/6$.
The decay rate of inflaton to the SM gauge bosons is then given by
\begin{equation}
    \Gamma_{\sigma \to AA} 
    =
    \sum_G \frac{b_G^2 \alpha_G^2 N_G}{768 \pi^3} \frac{m_\sigma^3}{\Mpl^2}~,
\end{equation}
where $N_{\text{SU}(3)_C} = 8$, $N_{\text{SU}(2)_L} = 3$, and $N_{\text{U}(1)_Y} = 1$.

The inflaton mostly decays into the Higgs and $\chi$ at $H \sim \Gamma_\sigma$ with the total decay width being $\Gamma_\sigma$.
The reheating temperature is estimated as
\begin{equation}
    T_\text{R} \simeq \qty( \frac{90}{\pi^2 g_* (T_\text{R})} )^{1/4} \sqrt{\Gamma_{\sigma} \Mpl}
    \simeq 5 \times 10^{9} \,\mathrm{GeV}\, \qty( \frac{m_\sigma}{3 \times 10^{13}\,\mathrm{GeV}} )^{3/2}.
\end{equation}
Unless we introduce a non-minimal coupling between the Higgs field and the Ricci curvature, this value is a generic prediction of this setup.
The corresponding $e$-folding number of inflation is
\begin{equation}
    N_{e} = 55.7 + 
    \frac{1}{3} {\text{ln}} \qty(\frac{T_{\text{R}}}{5 \times 10^9\,{\text{GeV}}})
    + \frac{1}{4}{\text{ln}} \Delta,
\end{equation}
where $T_{\text{R}}$ is the reheating temperature and $\Delta$ is the diluting factor which will be introduced later.
We take the pivot scale as $k_p/a_0=0.002\,{\rm{Mpc}}^{-1}$.
This is depicted in Fig.~\ref{fig:nsr}.
Here we allow a certain amount of matter domination after reheating, which is indeed a necessary ingredient to match the observation as we see in Sec.~\ref{sec:R2-DR}.

\subsection{Dark radiation problem and its solution}
\label{sec:R2-DR}

Immediately after the production, the Higgs pairs interact with other SM massless particles and make the SM thermal bath of the temperature $T_\text{R} \sim 10^9 \,\mathrm{GeV}$.
On the other hand, the $\chi$ can only interact with the SM particles via the Planck-suppressed operators, and does not have any strong enough interaction with the SM particles at this temperature.
Hence it is decoupled from the SM thermal bath. Its energy is fixed at $m_{\sigma}/2$ which is only red-shifted by the universe expansion. The massless $\chi$ causes too much abundance of the dark radiation (DR), since the branching ratio of the inflaton decay to the $\chi \chi$ channel is sizable $\sim 20$\% unless the Higgs fields have a non-minimal coupling to the Ricci curvature: $\xi_H \abs{H}^2 R$~\cite{Gorbunov:2013dqa}. The case of $\xi_H \abs{H}^2 R$ will be discussed in detail in the next Sec.~\ref{sec:h2R}.

Here we provide a solution to this problem within the minimal coupling because the necessary ingredients are already present in the model.
The excess in the effective relativistic degrees of freedom, $\Delta N_\text{eff}$, is given by~\cite{Garcia-Bellido:2012npk,Gorbunov:2013dqa}
\begin{equation}
    \Delta N_\text{eff} 
    = \frac{43}{7} \qty( \frac{10.75}{g_\ast (T_\text{R})} )^{1/3}
    \left.\frac{\rho_\chi}{\rho_\text{rad}}\right|_{T_\text{R}} \times \Delta
    \simeq \frac{2.86}{4 + \sum_\alpha x_\alpha (1 - x_\alpha)^{3/2} / 2} \times \Delta, \qquad x_\alpha \equiv \frac{4 M_\alpha^2}{m_\sigma^2},
\end{equation}
where the index $\alpha$ should be summed over the Majorana right-handed neutrinos which decays without dominating the universe.
Since the Planck result combined with baryon acoustic oscillation data requires $\Delta N_\text{eff} \lesssim 0.3$~\cite{Planck:2018vyg}, we need a dilution factor $\Delta \lesssim 0.4$ to alleviate this tension.

To solve the above problem we consider one of the right-handed neutrino called as $\Nd$ is long-lived so that its decay provides the additional entropy in the SM thermal bath (late-time entropy production).
The dilution factor of $\Delta \lesssim 0.4$ is easily realized by choosing the mass and Yukawa coupling constants of $\Nd$.
For a given mass of $\Nd$, one may estimate the temperature of the SM thermal bath when $\Nd$ comes to dominate the universe as
\begin{equation}
    T_{\Nd,\text{dom}} \simeq \frac{\xd^{3/2} \qty( 1 - \xd)^{3/2}}{8}\, T_\text{R}.
\end{equation}
When the $\Nd$ decay rate becomes comparable to the Hubble parameter, it decays into the SM particles, which occurs at
\begin{equation}
    T_{\Nd,\text{dec}} \simeq \qty( \frac{90}{\pi^2 g_\ast (T_{\Nd,\text{dec}})} )^{1/4} 
    \qty( \frac{\sum_\alpha\abs{\yd}^2 }{8 \pi} )^{1/2}
    \sqrt{\Md \Mpl}.
\end{equation}
The dilution factor by the $\Nd$ decay is estimated as
\begin{equation}
    \Delta 
    \simeq 
    \qty( \frac{T_{\Nd,\text{dec}}}{T_{\Nd,\text{dom}}} )^{4/3}
    \simeq 0.2 \times \qty( \frac{\Md}{10^{11}\,\mathrm{GeV}} )^{-10/3}
    \Bigg[\sum_\alpha \bigg( \frac{\abs{y_{1\alpha}}}{10^{-12}} \bigg)^2
    \Bigg]^{2/3},
\end{equation}
where we have assumed $g_\ast (T_{\Nd,\text{dec}}) \simeq g_\ast (T_{\Nd,\text{dom}})$ for simplicity.
The seesaw formula suggests that one of the eigenvalues of the light neutrino mass matrix to be almost vanishing because of the required small Yukawa coupling constants and heavy mass term for $\Nd$.

By the decay of $\Nd$, the DR of $\chi$ generated by the inflaton decay is diluted.
Still, the thermal production of $\chi$ from the SM plasma is possible.
Since $\chi$ can only interact with the SM via the inflaton mediation or other Planck-suppressed operators, the amount of the thermal production of $\chi$ can be safely neglected.

\subsection{Leptogenesis}
\label{sec:lg_dm}

Now we discuss the leptogenesis \cite{Fukugita:1986hr}. We assume $\Nl$ is lighter than $\Nh$ and show that the $\Nl$ decay produces a large enough lepton asymmetry even after the late time entropy production discussed in the previous subsection.

We consider the Majorana mass of $\Nl$ much larger than $T_\text{R}$.
In this case, the main production of $\Nl$ is the decay of the inflaton.
We assume that there is no degeneracy in the Majorana mass terms, \textit{i.e.}, $\Ml \ll \Mh$.\footnote{
    An alternative way to overcome the dilution factor would be to assume a slight degeneracy between the masses of $\Nl$ and $\Nh$.
    We will not pursue this possibility in this paper.
}
The number density of $\Nl$ at the reheating temperature can be estimated as
\begin{equation}
    \left. \frac{n_{\Nl}}{s} \right|_{\text{R}} \simeq \frac{\Gamma_{\sigma \to \Nl\Nl}}{\Gamma_{\sigma}}\, \frac{3}{2}\, \frac{T_\text{R}}{m_\sigma},
\end{equation}
where the entropy density is denoted by $s$.
Immediately after the production, $\Nl$ decays into the lepton and Higgs in a $CP$ violating way, which produces the lepton asymmetry~\cite{Fukugita:1986hr}.
As the $\Nl$ is much heavier than the temperature of the SM thermal bath, the produced lepton asymmetry is not washed-out, known as the non-thermal leptogenesis~\cite{Lazarides:1990huy,Kumekawa:1994gx,Asaka:1999yd,Buchmuller:2005eh}.
After the non-thermal leptogenesis, the $\Nd$ decay that is required to alleviate the sizable DR of $\chi$ dilutes the lepton asymmetry by the factor of $\Delta^{3/4}$, which results in \cite{Harvey:1990qw}
\begin{equation}
    \frac{n_B}{s}
    \simeq - \Delta^{3/4} \times \frac{28}{79}\, \epsilon_2\,  \left. \frac{n_{\Nl}}{s} \right|_{\text{R}}.
\end{equation}

The asymmetry parameter of the $\Nl$ decay can be expressed as~\cite{Buchmuller:2005eh}
\begin{equation}
    \epsilon_2
    \simeq
    - \frac{3}{16 \pi} \frac{\Im \big[ \big(y^\nu y^{\nu \dag}\big)_{23}^2 \big]}{ \big(y^\nu y^{\nu\dag}\big)_{22}} \frac{\Ml}{\Mh}
    \simeq
    - \frac{3}{16 \pi} \frac{\Ml m_{\nu_3}}{\langle \abs{H}^2 \rangle}\, \delta_\text{eff},
\end{equation}
where the heaviest active neutrino mass is denoted as $m_{\nu_3}$.
The effective $CP$ violation parameter is bounded by $|\delta_\text{eff}| \leqslant 1$, and thereby the asymmetry parameter $|\epsilon_2|$ has an upper bound~\cite{Hamaguchi:2001gw,Davidson:2002qv}.
Using these expressions, we find that the baryon-to-entropy ratio is estimated as
\begin{equation}
    \frac{n_B}{s} 
    \sim 10^{-10} \, \qty( \frac{\Delta}{0.1} )^{3/4}\, \qty( \frac{T_\text{R}}{5 \times 10^{9}\,\mathrm{GeV}} )\,
    \qty( \frac{\Ml}{0.2\, m_\sigma} )^3\,
    \qty( \frac{m_{\nu_3}}{0.05 \,\mathrm{eV}} )\,
    \delta_\text{eff},
\end{equation}
where we have omitted the factor $(1 - 4 \Ml^2/m_\sigma^2)^{3/2}$ for notational brevity, which is practically fine as long as $4 \Ml^2 / m_\sigma^2 \ll 1$.
Now we see that the observed baryon asymmetry in the present universe, $n_B/s \simeq 10^{-10}$, can be explained even if we have the required late-time dilution of $\Delta \lesssim 0.4$.

Note that, in the inverted hierarchy of the neutrino mass, the effective $CP$ violating parameter, $\delta_\text{eff}$, is suppressed by a factor of $\Delta m_\odot^2 / (2 \Delta m_\text{atm}^2) \simeq 0.015$~\cite{Davidson:2002qv,Harigaya:2012bw,Higaki:2014dwa} with the mass-squared difference for the solar and atmospheric neutrinos being $\Delta m_\odot^2 \simeq ( 0.0086  \, \mathrm{eV} )^2$ and $\Delta m_\text{atm}^2 \simeq (0.05 \,\mathrm{eV})^2$, respectively.
This scenario is already in tension with the observed baryon asymmetry even when it is maximized by setting $\Ml = m_\sigma/ \sqrt{8}$.
One might alleviate this discrepancy by assuming a slight degeneracy between the masses of $\Nl$ and $\Nh$.\footnote{
    \label{fn:htenrion?}
    One could further allow a certain deviation of $\Delta N_\text{eff} \sim 0.5$ motivated by the Hubble tension.
    
}

\section{Adding non-minimal coupling to Higgs}
\label{sec:h2R}

So far we only consider the case where the Higgs has a minimal coupling to the Ricci curvature for simplicity.
In this section, we also turn on the non-minimal coupling between the Higgs and the Ricci curvature, which can be regarded a more general theory consistent with our construction~\cite{Gorbunov:2013dqa}.\footnote{
    See also Refs.~\cite{Salvio:2014soa,Kannike:2015apa} for different but related setups.
}
The implications of the non-minimal coupling to the Higgs are two folds; it changes the inflationary trajectory and the reheating process.

Our defining action is given by
\begin{align}
    \mathcal{L} 
    &= \frac{1}{2} \qty( \xi \phi^2 + 2 \xi_H \abs{H}^2 ) R + \frac{1}{2} \qty(\partial \phi)^2 + \abs{\partial H}^2 
    - \frac{\lambda}{4} \phi^4
    - \frac{\lambda_H}{4} \qty( \abs{H}^2 - \lambda_v^2 \phi^2 )^2
    + \alpha R^2 + \mathcal{L}_\text{SM}' \\
    &\rightarrow  \qty( \frac{\xi}{2} \phi^2 + \xi_H \abs{H}^2 + 2 \alpha X ) R - \alpha X^2 + \frac{1}{2} \qty(\partial \phi)^2
    + \abs{\partial H}^2
    - \frac{\lambda}{4} \phi^4
    - \frac{\lambda_H}{4} \qty( \abs{H}^2 - \lambda_v^2 \phi^2 )^2
    + \mathcal{L}_\text{SM}',
    \label{eq:h2R-jordanX}
\end{align}
where the Higgs four-point coupling is denoted as $\lambda_H$, the coupling $\lambda_v$ is taken to be consistent with the VEV of the Higgs field in the Einstein frame, and the coupling $\lambda$ is taken to be consistent with the current cosmological constant.\footref{fn:cc_running}
The SM Lagrangian except for the Higgs kinetic term and potential is denoted by $\mathcal{L}_\text{SM}'$.

One may utilize the similar technique in the previous section to obtain the action in the Einstein frame, where the massive and massless degrees of freedom are manifest.
The Weyl transformation in this case is given by
\begin{equation}
    \label{eq:Weyl-h2R}
    \Omega^2 = \frac{\xi \phi^2 + \xi_H h^2 + 4 \alpha X}{\Mpl^2},
\end{equation}
with the radial mode of the Higgs being $h = (2 \abs{H}^2)^{1/2}$.
One may replace all the $X$ fields in the action by means of $\Omega$ instead.
We introduce the following field redefinitions from $(\Omega, \phi, h)$ to $(\Theta, \chi, \Phi )$:
\begin{equation}
    \label{eq:field-redef-h2R}
    \Omega^2 = \frac{1}{6\Theta} e^{ \sqrt{\frac{2}{3}} \frac{\chi}{\Mpl} }, \qquad
    \frac{\phi^2}{\Mpl^2} = \qty( 1 - \Theta^{-1} ) \cos^2 \Phi\, e^{ \sqrt{\frac{2}{3}} \frac{\chi}{\Mpl} }, \qquad
    \frac{h^2}{\Mpl^2} = \qty( 1 - \Theta^{-1} ) \sin^2 \Phi\, e^{ \sqrt{\frac{2}{3}} \frac{\chi}{\Mpl} }.
\end{equation}
The field range of $\Theta$ is $\Theta \geqslant 1$ because these definitions imply
\begin{equation}
    \Theta = \frac{\qty( \xi + 1/6 ) \phi^2 + \qty( \xi_H + 1/6 ) h^2 + 4 \alpha X}{\xi \phi^2 + \xi_H h^2 + 4 \alpha X}.
\end{equation}

Performing the Weyl transformation \eqref{eq:Weyl-h2R} and field redefinitions \eqref{eq:field-redef-h2R} to Eq.~\eqref{eq:h2R-jordanX}, we obtain the following form of the action that can be regarded as a counterpart of Eqs.~\eqref{eq:R2_kin} and \eqref{eq:R2_pot}:
\begin{align}
    \label{eq:h2R-kin}
    \mathcal{L} &= \frac{\Mpl^2}{2} R_\text{E} + \frac{1}{2} \Theta\, \qty(\partial \chi)^2 + \frac{3}{4} \Mpl^2\, \frac{1}{\Theta \qty( \Theta - 1 )}\, \qty(\partial \Theta)^2
    + \frac{1}{2} \cdot 6 \Mpl^2 \qty( \Theta - 1 ) \qty[ \qty( \partial \Phi )^2 + \sin^2 \Phi\, \qty( \partial \bm{\pi} )^2 ] \\
    \label{eq:h2R-pot}
    & \quad 
    - \frac{\Mpl^4}{16 \alpha} \qty{ 1 - 6 \qty( \Theta - 1 ) \qty[ \xi + \qty( \xi_H - \xi ) \sin^2 \Phi ] }^2
    - \frac{9}{4} \lambda_H \Mpl^4 \qty( \Theta - 1 )^2 \sin^4 \Phi 
    + \Omega^{-4} \mathcal{L}_{\text{SM}'}.
\end{align}
Here we have introduced the angular mode of the Higgs field as $\bm{\pi}$, which is restricted to a unit sphere, \textit{i.e.,} $\bm{\pi} \cdot \bm{\pi} = 1$.
We drop $\lambda$ and $\lambda_v$ as they are negligibly small compared to other parameters.

Recall that, in our construction, all the dimensionful scales including the renormalization scale are replaced with $\phi$, and hence the coupling to the SM fields in the Einstein frame is solely controlled by
\begin{equation}
    \label{eq:h2R-SM-coupling}
    \frac{\phi^2}{\Omega^2} = 6 \Mpl^2 \, \qty( \Theta - 1 )\, \cos^2 \Phi,
\end{equation}
except for the kinetic mixing with the Higgs field.
From Eqs.~\eqref{eq:h2R-kin}, \eqref{eq:h2R-pot}, and \eqref{eq:h2R-SM-coupling}, one may confirm that the $\chi$ field has the shift symmetry $\chi \to \chi + \text{const.}$ and is decoupled from the SM fields as in the previous sections.

\subsection{Inflation and reheating}
Apparently, the kinetic term of $\Theta$ is enhanced in the limit of $\Theta \to 1 +$ and the potential becomes flattened around that point, which is a suitable candidate of the slow-roll inflation.
Contrary to the previous example, now we have additional fields $\Phi$ and $\bm{\pi}$, which are originated from the Higgs field.
The Nambu--Goldstone modes $\bm{\pi}$ turn out to be irrelevant during inflation~\cite{Ema:2023dxm} and hence we drop them in the following discussion of inflation.\footnote{
    The Nambu--Goldstone modes $\bm{\pi}$ are the longitudinal modes of $\mathrm{SU}(2)$ gauge bosons in the broken phase of the SM, where $h^2 = 2\langle |H|^2 \rangle \neq 0$.
}
The question is whether the $\Phi$ field modifies the pole/potential structure of the $\Theta$ field.

As the Higgs four-point coupling is not so small $\lambda_H \sim 0.01$, the $\Phi$ field acquires significant stabilizing potential from the second term in Eq.~\eqref{eq:h2R-pot} while the first term in Eq.~\eqref{eq:h2R-pot} supports the non-vanishing VEV of $\Phi$.
Since the effect of the stabilizing potential is typically much larger than the inflation scale, one may assume that the $\Phi$ field always tracks the minimum of the potential in the background of the non-zero VEV of the inflaton field $\Theta$~\cite{Ema:2017rqn,He:2018gyf}.
Motivated by this observation, we first minimize the potential with respect to $\Phi$, $\partial V / \partial \Phi$, in the background of inflaton, which is solved as
\begin{equation}
    \label{eq:Phi-min}
    \sin^2 \Phi \simeq
    \begin{cases}
        0 &\text{for} \quad \xi_H \leqslant \xi, \\
        \frac{\xi_H - \xi}{\alpha \lambda_H + \qty( \xi_H - \xi )^2} \frac{1 - 6 \xi \qty( \Theta - 1 )}{6 \qty( \Theta - 1 )} &\text{for} \quad \xi_H > \xi,
    \end{cases}
\end{equation}
where we have dropped $\lambda$ and $\lambda_v$ as they are negligible.
As will be confirmed shortly, one finds $6 \xi (\Theta - 1) \ll 1$ during inflation and $\xi \ll 1$ for a successful inflation.
The previous example of the minimal coupling case $\xi_H = 0$ is included in the first case, where the $\Phi$ field is stabilized at the origin and it is almost the Higgs field as can be seen from Eq.~\eqref{eq:field-redef-h2R}.
Hence, for $\xi_H > \xi$, the mixing becomes significant, $\Phi \sim \mathcal{O}(1)$, in the deep inflationary regime, indicating that the inflationary trajectory must involve the Higgs field.
Notice that the case $\xi_H < \xi$ includes a negative non-minimal coupling of the Higgs, where the Higgs field is stabilized at the origin by the Hubble induced mass term.

Plugging this solution \eqref{eq:Phi-min} back into the potential \eqref{eq:h2R-pot}, we obtain the following form of the potential relevant for inflation:
\begin{equation}
    \label{eq:h2R-pot-infl}
    V(\Theta) \simeq \frac{\Mpl^4}{16 \alpha} 
    \qty[ 1 - 6 \xi \qty( \Theta - 1 ) ]^2
    \begin{cases}
        1 &\text{for} \quad \xi_H \leqslant \xi, \\
        \frac{\alpha \lambda_H}{\alpha \lambda_H + \qty( \xi_H - \xi )^2}
        &\text{for} \quad \xi_H > \xi.
    \end{cases}
\end{equation}
Also the kinetic term of $\Theta$ receives a certain correction from the $\Phi$ field, which is obtained by inserting Eq.~\eqref{eq:Phi-min} into the kinetic term of $\Theta$ in Eq.~\eqref{eq:h2R-kin}:
\begin{equation}
    \label{eq:h2R-kin-infl}
    \mathcal{L}_\text{kin} (\Theta) \simeq \frac{3}{4} \Mpl^2 \frac{\Theta_1}{\Theta \qty( \Theta - \Theta_1 )} \qty(\partial \Theta)^2, \qquad
    \Theta_1 \equiv
    \begin{cases}
        1 &\text{for} \quad \xi_H \leqslant \xi, \\
        1 + \frac{1}{6} \frac{\xi_H}{\alpha \lambda_H + \xi_H^2}
        &\text{for} \quad \xi_H > \xi,
    \end{cases}
\end{equation}
where we have used $\xi \ll 1$ and $\xi \ll \xi_H$ in the second line of $\Theta_1$.
Since if $\xi_H \ll 1$, the scenario will be almost the same as in the last section, we focus on $\xi_H \gtrsim 1$.
For a successful inflation, we need $\alpha \sim 10^8$ or $\xi_H \sim 10^4 \lambda_H^{1/2}$ to make the inflation scale consistent with the CMB observations.
Hence, as long as the Higgs four-point coupling $\lambda_H$ is not so suppressed, the location of the pole relevant for the inflation is approximated with $\Theta_1 \simeq 1$.
From this fact together with Eqs.~\eqref{eq:h2R-pot-infl} and \eqref{eq:h2R-kin-infl}, we confirm that the functional form of the inflationary Lagrangian remains almost the same as the previous case in Sec.~\ref{sec:inf_ref_R2}.
Therefore, the inflationary prediction is unaffected once we fix the inflation scale correctly to match the CMB observations.

It is instructive to see the relation of our model to the Higgs-Dilaton model~\cite{Shaposhnikov:2008xb,Garcia-Bellido:2011kqb} here.
By taking the limit of $\alpha \lambda_H \ll \xi_H^2$ in Eq.~\eqref{eq:Phi-min}, one may reproduce the trajectory of the Higgs-Dilaton inflation.
Also, this limit implies that the inflation energy is now dominated by the Higgs potential instead of the $R^2$ term as can be seen from $V(\Theta) / \Mpl^4 \to \lambda_H /(16 \xi_H^2)$ for $\Theta \to 1+$.
Since the Higgs field has sizable top and neutrino Yukawa couplings in our model, such a large Higgs VEV during inflation may pose a question on the quantum corrections to the Higgs potential and thereby inflaton potential.
Indeed, as shown in Ref.~\cite{Bezrukov:2012hx}, the Higgs-Dilaton inflation is not stable against the quantum corrections under the prescription of our interest where the renormalization scale is specified by $\phi$.\footnote{
    They instead adopted the renormalization scale including also the Higgs field, which solves this issue~\cite{Bezrukov:2012hx}.
}
For this reason, we mostly focus on the case where the non-minimal coupling to the Higgs is not so large, $\xi_H^2 \ll \alpha \lambda_H$, and do not consider the Higgs-Dilaton limit in detail.
In this case, the relevant quantum corrections are the Higgs loops to the coefficient of the $R^2$ term, which is expected to be negligible as long as $\alpha \gg \xi_H^2$.

Now we move on to the discussion on the reheating process.
Although the inflationary prediction is almost the same as the previous minimal coupling case, the reheating process is significantly modified by the non-minimal coupling to the Higgs as we will see in the following.
Let us expand the fields around the vacuum at $\langle\Phi \rangle = 0$ and $\langle \Theta \rangle \simeq 1 + 1 / (6 \xi)$.
The coupling to the SM fields around the vacuum can be expressed as
\begin{equation}
    \frac{\phi^2}{\Omega^2} \simeq
    \frac{\Mpl^2}{\xi}
    \qty(
        1 - \sqrt{ \frac{2}{3} } \frac{\sigma}{\Mpl} -2 \xi \frac{\abs{H}^2}{\Mpl^2}
    ),
\end{equation}
after canonically normalizing the Higgs.
We find the following potential for the inflaton and the Higgs field around the vacuum:
\begin{equation}
    \label{eq:h2R-pot-reh}
    V (\sigma, H) \simeq 
    \frac{\Mpl^4}{16 \alpha} \qty[
        - \sqrt{\frac{2}{3}} \frac{\sigma}{\Mpl}
        + 2 \qty( \xi_H - \xi ) \frac{|H|^2}{\Mpl^2}
    ]^2
    + \frac{\lambda_H}{4} \abs{H}^4,
\end{equation}
where we have redefined the $\sigma$ field so that it becomes $\sigma = 0$ at vacuum.
After inflation, the $\sigma$ field oscillates around its minimum with the amplitude $\sim \mathcal{O}(0.1) \Mpl$.
The oscillatory trajectory can be classified into two regimes depending on the competition between $\alpha \lambda_H$ and $\xi_H^2$.

If the Higgs non-minimal coupling is large enough to fulfill $\xi_H^2 \gg \alpha \lambda_H$, the first term dominates over the second term in Eq.~\eqref{eq:h2R-pot-reh}, and hence the Higgs field oscillation is triggered by the $\sigma$ oscillation to cancel out the first term in Eq.~\eqref{eq:h2R-pot-reh}:
\begin{equation}
    \abs{H}^2 \simeq \frac{\Mpl}{\sqrt{6}\,\xi_H} \sigma \quad \longrightarrow \quad
    \frac{\lambda_H}{4} \abs{H}^4 \simeq 
    \frac{\lambda_H \Mpl^2}{24 \xi_H^2}\, \sigma^2,
\end{equation}
which reproduces the inflaton oscillation after the Higgs inflation.
Since the Higgs field couples to all the SM fields including the Majorana right-handed neutrinos, the preheating/reheating process is expected to be efficient.
However, for $\xi_H^2 \gg \alpha \lambda_H$, the NG boson $\pi$s or equivalently the longitudinal modes of the SM gauge bosons are produced violently, whose energy exceeds the perturbative unitarity bound~\cite{Ema:2016dny,Sfakianakis:2018lzf}.
If the scalaron is light enough, \textit{i.e.,} $\xi_H^2 \lesssim \alpha \lambda_H$, this production is suppressed, which heals the issue of the Higgs inflation~\cite{Gorbunov:2018llf,He:2018mgb,Ema:2019fdd,Ema:2020zvg}.
Under this condition, the reheating temperature is expected to be similar to the case of the Higgs-$R^2$ inflation~\cite{He:2018mgb,Bezrukov:2019ylq,He:2020qcb}.
We do not pursue this direction further in this paper.

On the other hand, if the Higgs non-minimal coupling is not so large, $\xi_H^2 \ll \alpha \lambda_H$, the Higgs field is still around the origin while the inflaton $\sigma$ oscillates around its minimum.
The reheating process is expected to be similar to the previous minimal coupling case, where the final stage of reheating is dominated by the perturbative inflaton decay into the Higgs, the Majorana right-handed neutrinos, and the $\chi$ field.
Still, the non-minimal coupling of the Higgs to the Ricci curvature changes the inflaton decay rate to the Higgs field.
On top of the kinetic term interaction~\eqref{eq:int_sigma_H}, we have a tri-linear interaction from Eq.~\eqref{eq:h2R-pot-reh}.
By using partial integrations and the equations of motion, one may simplify these interaction terms as follows:
\begin{equation}
    \mathcal{L}_{\sigma H} \simeq \qty( 6 \xi_H + 1 )\, \frac{m_\sigma^2}{2} \sqrt{\frac{2}{3}} \frac{\sigma}{\Mpl} \, \abs{H}^2,
\end{equation}
where we have dropped $\xi$ as it is much smaller than unity.
As the coupling is now enhanced by a factor of $6 \xi_H + 1$, the decay rate of inflaton to Higgs becomes
\begin{equation}
    \label{eq:h2R_Gamma_sigma_H}
    \Gamma_{\sigma \to H H^\ast} \simeq \qty( 6 \xi_H + 1 )^2 \frac{m_\sigma^3}{48 \pi \Mpl^2}.
\end{equation}
Therefore, one finds that the reheating temperature is significantly enhanced by just allowing an order one non-minimal coupling $\xi_H$
\begin{equation}
    T_\text{R} \simeq 3 \times 10^{10} \,\mathrm{GeV}\,
    \abs{ \xi_H + 1/6 }\,
    \qty( \frac{m_\sigma}{3 \times 10^{13}\,\mathrm{GeV}} )^{3/2}.
    \label{eq:TR_sec4}
\end{equation}
Here we have assumed $4 (6 \xi_H + 1)^2 > 1$, which is fulfilled unless the non-minimal coupling is close to the conformal one.

\subsection{Dark radiation, leptogenesis, and dark matter}
As we have seen in the previous subsection, the branching ratio of inflaton decay to the Higgs field is enhanced by a factor of $(6 \xi_H + 1)^2$.
This has a significant impact on the DR of $\chi$ as the amount of $\chi$ is reduced as~\cite{Gorbunov:2013dqa}
\begin{equation}
    \label{eq:h2R_Neff}
    \Delta N_\text{eff} \simeq \frac{43}{7}  \qty( \frac{10.75}{g_\ast (T_\text{R})} )^{1/3} \frac{1}{4 \qty( 6 \xi_H + 1 )^2} \simeq 0.02 \, \qty( \xi_H + 1/6 )^{-2},
\end{equation}
where we have assumed $4 (6 \xi_H + 1)^2 > 1$.
An order one $\xi_H$ is sufficient to alleviate the dark radiation problem, without additional dilution factor.

The absence of the late-time entropy production and the enhancement of the reheating temperature open up the following simple scenario for the leptogenesis and the dark matter, where we only have three Majorana right-handed neutrinos, $\Nd$, $\Nl$, and $\Nh$.
As in the previous Sec.~\ref{sec:lg_dm}, we assume that the decay of $\Nl$ produces a lepton asymmetry.
Now, the thermal production of $\Nl$ is sufficient, leading to the standard result based on the thermal leptogenesis:
\begin{equation}
    \frac{n_B}{s} \simeq - \kappa \times \frac{28}{79}\,
    \epsilon_2\, 
    \frac{n_{\Nl}^\text{(eq)}}{s}
    \simeq 10^{-10} \, \bigg( \frac{\kappa}{0.1} \bigg)
    \bigg( \frac{\Ml}{10^{10}\,\mathrm{GeV}} \bigg)
    \bigg( \frac{m_{\nu_3}}{0.05 \,\mathrm{eV}} \bigg)\,
    \delta_\text{eff},
\end{equation}
with the wash-out factor being $\kappa$.
Again, an order one $\xi_H$ is sufficient to explain the observed baryon asymmetry fulfilling the condition for the thermal production, $T_\text{R} > \Ml$.
Besides, non-thermal leptogenesis similar to that in the last section is indeed also possible.

In addition, one may assume that the lightest Majorana right-handed neutrino $\Nd$ is sufficiently long-lived to be the DM.
The production of $\Nd$ is dominated by the inflaton decay, which leads to
\begin{equation}
    \Omega_{\Nd} h^2 \simeq 0.16 \,
    \abs{ \xi_H + \frac{1}{6} }^{-1}
    \bigg( \frac{\Md}{10^{-6}\, m_\sigma} \bigg)^3
    \bigg( \frac{m_\sigma}{3 \times 10^{13}\,\mathrm{GeV}} \bigg)^{3/2},
\end{equation}
for $4 (6 \xi_H + 1)^2 > 1$.
In order to fulfill the stability of the DM, its neutrino Yukawa coupling must be extremely suppressed, or instead is forbidden by a $\mathbb{Z}_2$ symmetry where only the $\Nd$ is odd under the $\mathbb{Z}_2$.
In either case, the seesaw formula suggests that one of the active neutrino mass is (almost) vanishing.

As an extreme example, let us briefly consider the special case where the Higgs has a (nearly) conformal coupling to the Ricci curvature, $\xi_H \simeq -1/6$.
In this case, the dominant decay channel of the inflaton becomes the $\chi$ field, and the SM particles are only generated through the heavy Majorana right-handed neutrinos or the gauge fields, depending on the size of the Majorana mass.
Hence, the DR problem of $\chi$ is severe in this case.
We take $\Ml \simeq m_\sigma / \sqrt{10}$ to maximize the branching ratio of $\Nl$.
The excess in the effective number of neutrinos is estimated as $\Delta N_\text{eff} \simeq 30 \, \Delta$, which calls for a significant dilution factor $\Delta \lesssim 0.01$.
As discussed in the previous Sec.~\ref{sec:inf_ref_R2}, such a dilution factor is easily achieved by the late-time decay of $\Nd$ by an appropriate choice of the Majorana mass $\Md$ and the neutrino Yukawa $\yd$.
A non-trivial question is whether we can still explain the observed baryon asymmetry in the presence of such a significant dilution factor within this scenario.
As in the previous Sec.~\ref{sec:lg_dm}, we consider the non-thermal leptogenesis by the decay of $\Nl$.
Although this is a corner of the parameter space, we can still realize $n_B / s \simeq 10^{-10}$ consistent with the observed value.

Finally, we comment on the stability of the electroweak vacuum, and its relation to $\Delta N_\text{eff}$.
It is known that the electroweak vacuum is metastable within the SM, if there exists no new physics that stabilizes the Higgs potential below the instability scale $\sim 10^{11} \,\mathrm{GeV}$.
If this is really the case, the high-scale inflation, such as the Starobinsky inflation, is in severe tension with the stability of our vacuum because the Higgs field acquires fluctuations of the order of $H_\text{inf}$ that is much larger than the instability scale~\cite{Kobakhidze:2013tn,Fairbairn:2014zia,Hook:2014uia}.
This situation can be ameliorated by the presence of the negative non-minimal coupling to the Ricci curvature, which stabilizes the Higgs potential by the positive Hubble induced mass during inflation as long as $\xi_H \lesssim - 0.1$~\cite{Espinosa:2007qp,Kamada:2014ufa,Espinosa:2015qea} [see also Eq.~\eqref{eq:Phi-min}].
Yet, the effect of the non-minimal coupling is in fact a double-edged sword, as the Ricci scalar oscillation after inflation in turn enhances the Higgs fluctuations via the tachyonic instability~\cite{Herranen:2014cua}.
To avoid the decay of our vacuum by this effect, we have the following necessary condition, $- 1.7$\,-\,$-1.6 \lesssim \xi_H$~\cite{Ema:2016kpf,Figueroa:2017slm,Li:2022ugn}.
As we have demonstrated already, such an order one non-minimal coupling is perfectly consistent with the observed baryon asymmetry and the DM abundance, as well as the DR problem.
Moreover, the stability of our vacuum in this scenario implies $\Delta N_\text{eff} \gtrsim 0.01$ from Eq.~\eqref{eq:h2R_Neff}.
Note that, given that the lower bound on $\xi_H$ is a necessary condition, the actual bound might be more stringent.\footnote{
    For our lives, we need to make sure that all the horizons after inflation which are now causal, \textit{i.e.,} $\text{\# of horizons} \sim e^{3 N}$, should not exhibit the vacuum decay.
    This consideration might put a more stringent bound on $\xi_H$.
}

\section{Conclusions and Discussion}
The Planck scale $\Mpl$ may not be a fundamental constant.
The \textit{no-scale BD gravity} is a concrete realization of this idea by replacing all the dimensionful scales, which even include the scale associated with the renormalization, with the scalar field $\phi$.
As a consequence, the no-scale BD gravity has quantum scale invariance, and thereby
we have a massless scalar boson together with the massless graviton. 
Unlike in original BD gravity, there is no long range force caused by the scalar boson exchange.  Thus, the no-scale BD gravity is consistent with all observations even if the scalar boson is massless or ultralight.

Then we have turned on the $R^2$ term as there is no reason to suppress it in our construction.
This model turns out to have an additional massive scalar boson whose potential can be very similar to the Starobinsky inflaton potential.
We have checked our inflaton potential is consistent with all observations as the Starobinsky inflation model.
However, a potential pitfall of this scenario is that the branching ratio for the inflaton decay to the $\chi +\chi$ channel is relatively large, which leads to a too much DR at the present.
We have shown that the late-time decay of the right-handed neutrino $\Nd$ can solve this problem, while the non-thermal leptogenesis of $\Nl$ decay gives rise to the observed baryon asymmetry in the universe.

A more promising possibility is to introduce a non-minimal coupling between the Higgs and the Ricci curvature, $\xi_H \abs{H}^2 R$, which is another term allowed in our construction.
Owing to the coupling between the inflaton and the Higgs induced by the presence of $\xi_H$, the branching ratio for the inflaton decay to the $\chi +\chi$ channel is relatively suppressed.
We have shown that an order one $\xi_H$ is sufficient to solve the DR problem, whose value is sufficiently small that the inflationary prediction of the $R^2$ term is not affected.
This order one $\xi_H$ also increases the reheating temperature to $T_R\simeq 3\times10^{10}$ GeV, where the thermal leptogenesis by the decay of $\Nl$ is successful (since the late-time entropy production is not needed to solve the DR problem in this case).
As for the DM, we have identified it with one of the right-handed neutrinos, $\Nd$, assuming its life-time is sufficiently longer than the age of the universe. Since its Yukawa coupling is very small, one of the neutrino masses is predicted to be almost vanishing. The mass of the DM is predicted most likely at $\sim 10$ PeV, whose decay might be seen in the IceCube experiments.

There is another bonus of the non-minimal coupling to the Higgs, that is, the stability of the electroweak vacuum.
The stability of our vacuum during and after inflation implies the necessary condition of $-1.7$\,-\,$-1.6 \lesssim \xi_H \lesssim - 0.1$.
Such an order one $\xi_H$ is not only consistent with the observed baryon asymmetry and the DM abundance, but also with the DR constraints.
Interestingly, this necessary condition puts a lower bound of $\Delta N_\text{eff} \gtrsim 0.01$.
Provided that this lower bound on $\xi_H$ is merely a necessary condition, the actual bound might be more stringent, and hence the $\Delta N_\text{eff}$ can be an interesting observable to explore our scenario~\cite{Abazajian:2019eic}.

The massless-ness of the $\chi$ boson is guaranteed by the shift symmetry $\chi \rightarrow \chi + \text{const.}$ in the Einstein frame. This shift symmetry is not broken by any perturbative quantum corrections.
Still, it might be broken by non-perturbative effects in quantum gravity and a small mass term might be generated, that is, $\delta \mathcal{L}_\text{QG} = - (m^2/2) \chi^2$. If the mass is extremely small as $m\simeq 10^{-33}$ eV, the $\chi$ plays a role of the quintessence whose potential energy density contributes to the observed dark energy (DE). And if the mass is around $10^{-21}$ eV the $\chi$ can be the fuzzy DM. An interesting feature of the $\chi$ DM is ``no-coupling'' to the SM particles except for gravity at low energies, which might be tested in future observations.

However, on the other hand, it seems impossible to generate the corresponding potential term for the $\phi$ boson in the Jordan frame by any quantum corrections including the gravity, since there is no mass parameter in our no-scale theory. Thus, we can naively consider that any potentials including the mass terms for the $\chi$ boson in the Einstein frame are vanishing  even at the quantum-theory level, which suggests that the contributions from  non-perturbative effects in the quantum gravity are suppressed. This serious problem will be discussed in future publications.


An intriguing possibility to introduce an explicit breaking of the scale invariance is to introduce an constant $\Lambda_0$ in the Jordan frame, that is, $\mathcal{L}_\text{J}= - \Lambda_0$.
In the Einstein frame we have a potential for the $\chi$, that is, $\mathcal{L}_\text{E} = \Lambda_0 e^{- 4A \chi/\Mpl}$ where $A = \sqrt{1/6} \sqrt{6\xi/(6 \xi + 1)}$. Here, we choose the origin of the $\chi$ so that the dark energy is exactly the same as $\Lambda_0$. If the $\chi$ is now starting to move, we may test this potential by future observations.

Another interesting effect that has not been discussed in this paper is higher-dimensional operators.
The BD theory is not well defined at the origin of the $\phi$, since the gravitational coupling is infinity at the origin. 
Hence we may suppose that the $\phi=0$ point should be excluded in the theory.
In this case there is no reason to forbid higher-dimensional operators such as $R^3/\phi^2+ \cdots $, which are consistent with the scale invariance.
Inclusion of such higher order terms makes the theory is defined although we have infinite number of the higher dimensional operators.
Here, the field $\phi$ plays as a cut-off scale which becomes $M_\text{Pl}$ in the Einstein frame as explained in Sec.~\ref{sec:no-scale-BD}.
Possible phenomenological implications of these higher-dimensional operators will be discussed in future publications.

\section*{Acknowledgment}
We thank Qiuyue Liang for discussion on the Brans-Dicke gravity.
M.\,H.~is supported by Grant-in-Aid for JSPS Fellows 23KJ0697.
K.\,M.~is supported by JSPS KAKENHI Grant No.~JP22K14044.
T.\,T.\,Y.~is supported by the Natural Science Foundation of China (NSFC) under Grant No.~12175134, MEXT KAKENHI Grants No.~JP24H02244, and
World Premier International Research Center Initiative
(WPI Initiative), MEXT, Japan.

\small
\bibliographystyle{utphys}
\bibliography{ref}

@article{Cox:2017rgn,
    author = "Cox, Peter and Han, Chengcheng and Yanagida, Tsutomu T.",
    title = "{Right-handed Neutrino Dark Matter in a U(1) Extension of the Standard Model}",
    eprint = "1710.01585",
    archivePrefix = "arXiv",
    primaryClass = "hep-ph",
    reportNumber = "IPMU17-0138",
    doi = "10.1088/1475-7516/2018/01/029",
    journal = "JCAP",
    volume = "01",
    pages = "029",
    year = "2018"
}

@article{Kusenko:2010ik,
    author = "Kusenko, Alexander and Takahashi, Fuminobu and Yanagida, Tsutomu T.",
    title = "{Dark Matter from Split Seesaw}",
    eprint = "1006.1731",
    archivePrefix = "arXiv",
    primaryClass = "hep-ph",
    reportNumber = "IPMU-10-0095",
    doi = "10.1016/j.physletb.2010.08.031",
    journal = "Phys. Lett. B",
    volume = "693",
    pages = "144--148",
    year = "2010"
}

@article{Brans:1961sx,
    author = "Brans, C. and Dicke, R. H.",
    editor = "Hsu, Jong-Ping and Fine, D.",
    title = "{Mach's principle and a relativistic theory of gravitation}",
    doi = "10.1103/PhysRev.124.925",
    journal = "Phys. Rev.",
    volume = "124",
    pages = "925--935",
    year = "1961"
}

@article{Will:2005va,
    author = "Will, Clifford M.",
    title = "{The Confrontation between general relativity and experiment}",
    eprint = "gr-qc/0510072",
    archivePrefix = "arXiv",
    doi = "10.12942/lrr-2006-3",
    journal = "Living Rev. Rel.",
    volume = "9",
    pages = "3",
    year = "2006"
}

@article{Ferreira:2016kxi,
    author = "Ferreira, Pedro G. and Hill, Christopher T. and Ross, Graham G.",
    title = "{No fifth force in a scale invariant universe}",
    eprint = "1612.03157",
    archivePrefix = "arXiv",
    primaryClass = "gr-qc",
    reportNumber = "FERMILAB-PUB-16-665-T",
    doi = "10.1103/PhysRevD.95.064038",
    journal = "Phys. Rev. D",
    volume = "95",
    number = "6",
    pages = "064038",
    year = "2017"
}

@article{Rinaldi:2015uvu,
    author = "Rinaldi, Massimiliano and Vanzo, Luciano",
    title = "{Inflation and reheating in theories with spontaneous scale invariance symmetry breaking}",
    eprint = "1512.07186",
    archivePrefix = "arXiv",
    primaryClass = "gr-qc",
    doi = "10.1103/PhysRevD.94.024009",
    journal = "Phys. Rev. D",
    volume = "94",
    number = "2",
    pages = "024009",
    year = "2016"
}

@article{Tambalo:2016eqr,
    author = "Tambalo, Giovanni and Rinaldi, Massimiliano",
    title = "{Inflation and reheating in scale-invariant scalar-tensor gravity}",
    eprint = "1610.06478",
    archivePrefix = "arXiv",
    primaryClass = "gr-qc",
    doi = "10.1007/s10714-017-2217-8",
    journal = "Gen. Rel. Grav.",
    volume = "49",
    number = "4",
    pages = "52",
    year = "2017"
}

@article{Fukugita:1986hr,
    author = "Fukugita, M. and Yanagida, T.",
    title = "{Baryogenesis Without Grand Unification}",
    reportNumber = "RIFP-641",
    doi = "10.1016/0370-2693(86)91126-3",
    journal = "Phys. Lett. B",
    volume = "174",
    pages = "45--47",
    year = "1986"
}

@article{Buchmuller:2005eh,
    author = "Buchmuller, W. and Peccei, R. D. and Yanagida, T.",
    title = "{Leptogenesis as the origin of matter}",
    eprint = "hep-ph/0502169",
    archivePrefix = "arXiv",
    reportNumber = "DESY-05-031",
    doi = "10.1146/annurev.nucl.55.090704.151558",
    journal = "Ann. Rev. Nucl. Part. Sci.",
    volume = "55",
    pages = "311--355",
    year = "2005"
}

@article{Burrage:2018dvt,
    author = "Burrage, Clare and Copeland, Edmund J. and Millington, Peter and Spannowsky, Michael",
    title = "{Fifth forces, Higgs portals and broken scale invariance}",
    eprint = "1804.07180",
    archivePrefix = "arXiv",
    primaryClass = "hep-th",
    reportNumber = "IPPP/18/23, IPPP-18-23",
    doi = "10.1088/1475-7516/2018/11/036",
    journal = "JCAP",
    volume = "11",
    pages = "036",
    year = "2018"
}

@article{Englert:1976ep,
    author = "Englert, F. and Truffin, C. and Gastmans, R.",
    title = "{Conformal Invariance in Quantum Gravity}",
    reportNumber = "PRINT-76-0296 (BRUSSELS)",
    doi = "10.1016/0550-3213(76)90406-5",
    journal = "Nucl. Phys. B",
    volume = "117",
    pages = "407--432",
    year = "1976"
}

@article{Shaposhnikov:2008xi,
    author = "Shaposhnikov, Mikhail and Zenhausern, Daniel",
    title = "{Quantum scale invariance, cosmological constant and hierarchy problem}",
    eprint = "0809.3406",
    archivePrefix = "arXiv",
    primaryClass = "hep-th",
    doi = "10.1016/j.physletb.2008.11.041",
    journal = "Phys. Lett. B",
    volume = "671",
    pages = "162--166",
    year = "2009"
}

@article{Starobinsky:1980te,
    author = "Starobinsky, Alexei A.",
    editor = "Khalatnikov, I. M. and Mineev, V. P.",
    title = "{A New Type of Isotropic Cosmological Models Without Singularity}",
    doi = "10.1016/0370-2693(80)90670-X",
    journal = "Phys. Lett. B",
    volume = "91",
    pages = "99--102",
    year = "1980"
}

@article{Vilenkin:1985md,
    author = "Vilenkin, Alexander",
    title = "{Classical and Quantum Cosmology of the Starobinsky Inflationary Model}",
    reportNumber = "HUTP-85-A017",
    doi = "10.1103/PhysRevD.32.2511",
    journal = "Phys. Rev. D",
    volume = "32",
    pages = "2511",
    year = "1985"
}

@article{Maeda:1987xf,
    author = "Maeda, Kei-ichi",
    title = "{Inflation as a Transient Attractor in R**2 Cosmology}",
    reportNumber = "UTAP-60-87",
    doi = "10.1103/PhysRevD.37.858",
    journal = "Phys. Rev. D",
    volume = "37",
    pages = "858",
    year = "1988"
}

@article{Fukuda:1974kn,
    author = "Fukuda, Takashi and Kawasaki, Mamoru and Yanagida, Tsutomu and Yonezawa, Minoru",
    title = "{Pi eta Mixing}",
    reportNumber = "RRK 74-4",
    doi = "10.1143/PTP.53.1135",
    journal = "Prog. Theor. Phys.",
    volume = "53",
    pages = "1135",
    year = "1975"
}

@article{Armillis:2013wya,
    author = "Armillis, Roberta and Monin, Alexander and Shaposhnikov, Mikhail",
    title = "{Spontaneously Broken Conformal Symmetry: Dealing with the Trace Anomaly}",
    eprint = "1302.5619",
    archivePrefix = "arXiv",
    primaryClass = "hep-th",
    doi = "10.1007/JHEP10(2013)030",
    journal = "JHEP",
    volume = "10",
    pages = "030",
    year = "2013"
}

@article{Falls:2018olk,
    author = "Falls, Kevin and Herrero-Valea, Mario",
    title = "{Frame (In)equivalence in Quantum Field Theory and Cosmology}",
    eprint = "1812.08187",
    archivePrefix = "arXiv",
    primaryClass = "hep-th",
    doi = "10.1140/epjc/s10052-019-7070-3",
    journal = "Eur. Phys. J. C",
    volume = "79",
    number = "7",
    pages = "595",
    year = "2019"
}

@article{Jeong:2023zrv,
    author = "Jeong, Hyun and Kamada, Kohei and Starobinsky, Alexei A. and Yokoyama, Jun'ichi",
    title = "{Reheating process in the R $^{2}$ inflationary model with the baryogenesis scenario}",
    eprint = "2305.14273",
    archivePrefix = "arXiv",
    primaryClass = "hep-ph",
    reportNumber = "RESCEU-13/23",
    doi = "10.1088/1475-7516/2023/11/023",
    journal = "JCAP",
    volume = "11",
    pages = "023",
    year = "2023"
}

@article{Gorbunov:2013dqa,
    author = "Gorbunov, Dmitry and Tokareva, Anna",
    title = "{Scale-invariance as the origin of dark radiation?}",
    eprint = "1307.5298",
    archivePrefix = "arXiv",
    primaryClass = "astro-ph.CO",
    doi = "10.1016/j.physletb.2014.10.036",
    journal = "Phys. Lett. B",
    volume = "739",
    pages = "50--55",
    year = "2014"
}

@article{Garcia-Bellido:2012npk,
    author = "Garcia-Bellido, Juan and Rubio, Javier and Shaposhnikov, Mikhail",
    title = "{Higgs-Dilaton cosmology: Are there extra relativistic species?}",
    eprint = "1209.2119",
    archivePrefix = "arXiv",
    primaryClass = "hep-ph",
    reportNumber = "PREPRINT-IFT-UAM-CSIC-12-87",
    doi = "10.1016/j.physletb.2012.10.075",
    journal = "Phys. Lett. B",
    volume = "718",
    pages = "507--511",
    year = "2012"
}

@article{Planck:2018vyg,
    author = "Aghanim, N. and others",
    collaboration = "Planck",
    title = "{Planck 2018 results. VI. Cosmological parameters}",
    eprint = "1807.06209",
    archivePrefix = "arXiv",
    primaryClass = "astro-ph.CO",
    doi = "10.1051/0004-6361/201833910",
    journal = "Astron. Astrophys.",
    volume = "641",
    pages = "A6",
    year = "2020",
    note = "[Erratum: Astron.Astrophys. 652, C4 (2021)]"
}

@article{Lazarides:1990huy,
    author = "Lazarides, George and Shafi, Q.",
    title = "{Origin of matter in the inflationary cosmology}",
    reportNumber = "BA-90-78",
    doi = "10.1016/0370-2693(91)91090-I",
    journal = "Phys. Lett. B",
    volume = "258",
    pages = "305--309",
    year = "1991"
}

@article{Kumekawa:1994gx,
    author = "Kumekawa, Kazuya and Moroi, Takeo and Yanagida, Tsutomu",
    title = "{Flat potential for inflaton with a discrete R invariance in supergravity}",
    eprint = "hep-ph/9405337",
    archivePrefix = "arXiv",
    reportNumber = "TU-458",
    doi = "10.1143/PTP.92.437",
    journal = "Prog. Theor. Phys.",
    volume = "92",
    pages = "437--448",
    year = "1994"
}

@article{Asaka:1999yd,
    author = "Asaka, T. and Hamaguchi, Koichi and Kawasaki, M. and Yanagida, T.",
    title = "{Leptogenesis in inflaton decay}",
    eprint = "hep-ph/9906366",
    archivePrefix = "arXiv",
    reportNumber = "UT-853, RESCEU-15-99",
    doi = "10.1016/S0370-2693(99)01020-5",
    journal = "Phys. Lett. B",
    volume = "464",
    pages = "12--18",
    year = "1999"
}

@article{Hamaguchi:2001gw,
    author = "Hamaguchi, Koichi and Murayama, Hitoshi and Yanagida, T.",
    title = "{Leptogenesis from N dominated early universe}",
    eprint = "hep-ph/0109030",
    archivePrefix = "arXiv",
    reportNumber = "UT-957, LBNL-48679, UCB-PTH-01-30",
    doi = "10.1103/PhysRevD.65.043512",
    journal = "Phys. Rev. D",
    volume = "65",
    pages = "043512",
    year = "2002"
}

@article{Davidson:2002qv,
    author = "Davidson, Sacha and Ibarra, Alejandro",
    title = "{A Lower bound on the right-handed neutrino mass from leptogenesis}",
    eprint = "hep-ph/0202239",
    archivePrefix = "arXiv",
    reportNumber = "OUTP-02-10P, IPPP-02-16, DCPT-02-32",
    doi = "10.1016/S0370-2693(02)01735-5",
    journal = "Phys. Lett. B",
    volume = "535",
    pages = "25--32",
    year = "2002"
}

@article{Harigaya:2012bw,
    author = "Harigaya, Keisuke and Ibe, Masahiro and Yanagida, Tsutomu T.",
    title = "{Seesaw Mechanism with Occam's Razor}",
    eprint = "1205.2198",
    archivePrefix = "arXiv",
    primaryClass = "hep-ph",
    doi = "10.1103/PhysRevD.86.013002",
    journal = "Phys. Rev. D",
    volume = "86",
    pages = "013002",
    year = "2012"
}

@article{Higaki:2014dwa,
    author = "Higaki, Tetsutaro and Kitano, Ryuichiro and Sato, Ryosuke",
    title = "{Neutrinoful Universe}",
    eprint = "1405.0013",
    archivePrefix = "arXiv",
    primaryClass = "hep-ph",
    reportNumber = "KEK-TH-1731",
    doi = "10.1007/JHEP07(2014)044",
    journal = "JHEP",
    volume = "07",
    pages = "044",
    year = "2014"
}

@article{Ema:2023dxm,
    author = "Ema, Yohei and Verner, Sarunas",
    title = "{Cosmological collider signatures of Higgs-R$^{2}$ inflation}",
    eprint = "2309.10841",
    archivePrefix = "arXiv",
    primaryClass = "hep-ph",
    reportNumber = "UMN-TH-4224/23, FTPI-MINN-23-16",
    doi = "10.1088/1475-7516/2024/04/039",
    journal = "JCAP",
    volume = "04",
    pages = "039",
    year = "2024"
}

@article{Shaposhnikov:2008xb,
    author = "Shaposhnikov, Mikhail and Zenhausern, Daniel",
    title = "{Scale invariance, unimodular gravity and dark energy}",
    eprint = "0809.3395",
    archivePrefix = "arXiv",
    primaryClass = "hep-th",
    doi = "10.1016/j.physletb.2008.11.054",
    journal = "Phys. Lett. B",
    volume = "671",
    pages = "187--192",
    year = "2009"
}

@article{Garcia-Bellido:2011kqb,
    author = "Garcia-Bellido, Juan and Rubio, Javier and Shaposhnikov, Mikhail and Zenhausern, Daniel",
    title = "{Higgs-Dilaton Cosmology: From the Early to the Late Universe}",
    eprint = "1107.2163",
    archivePrefix = "arXiv",
    primaryClass = "hep-ph",
    reportNumber = "IFT-UAM-CSIC-11-49",
    doi = "10.1103/PhysRevD.84.123504",
    journal = "Phys. Rev. D",
    volume = "84",
    pages = "123504",
    year = "2011"
}

@article{Bezrukov:2012hx,
    author = "Bezrukov, Fedor and Karananas, Georgios K. and Rubio, Javier and Shaposhnikov, Mikhail",
    title = "{Higgs-Dilaton Cosmology: an effective field theory approach}",
    eprint = "1212.4148",
    archivePrefix = "arXiv",
    primaryClass = "hep-ph",
    reportNumber = "RBRC-1007",
    doi = "10.1103/PhysRevD.87.096001",
    journal = "Phys. Rev. D",
    volume = "87",
    number = "9",
    pages = "096001",
    year = "2013"
}

@article{He:2018mgb,
    author = "He, Minxi and Jinno, Ryusuke and Kamada, Kohei and Park, Seong Chan and Starobinsky, Alexei A. and Yokoyama, Jun'ichi",
    title = "{On the violent preheating in the mixed Higgs-$R^2$ inflationary model}",
    eprint = "1812.10099",
    archivePrefix = "arXiv",
    primaryClass = "hep-ph",
    reportNumber = "CTPU-PTC-18-43, LDU-18-007, RESCEU-16/18",
    doi = "10.1016/j.physletb.2019.02.008",
    journal = "Phys. Lett. B",
    volume = "791",
    pages = "36--42",
    year = "2019"
}

@article{Bezrukov:2019ylq,
    author = "Bezrukov, Fedor and Gorbunov, Dmitry and Shepherd, Chris and Tokareva, Anna",
    title = "{Some like it hot: $R^2$ heals Higgs inflation, but does not cool it}",
    eprint = "1904.04737",
    archivePrefix = "arXiv",
    primaryClass = "hep-ph",
    reportNumber = "INR-TH-2019-006, MAN/HEP/2019/001",
    doi = "10.1016/j.physletb.2019.06.064",
    journal = "Phys. Lett. B",
    volume = "795",
    pages = "657--665",
    year = "2019"
}

@article{He:2020qcb,
    author = "He, Minxi",
    title = "{Perturbative Reheating in the Mixed Higgs-$R^2$ Model}",
    eprint = "2010.11717",
    archivePrefix = "arXiv",
    primaryClass = "hep-ph",
    reportNumber = "RESCEU-20/20",
    doi = "10.1088/1475-7516/2021/05/021",
    journal = "JCAP",
    volume = "05",
    pages = "021",
    year = "2021"
}

@article{Casas:2017wjh,
    author = "Casas, Santiago and Pauly, Martin and Rubio, Javier",
    title = "{Higgs-dilaton cosmology: An inflation\textendash{}dark-energy connection and forecasts for future galaxy surveys}",
    eprint = "1712.04956",
    archivePrefix = "arXiv",
    primaryClass = "astro-ph.CO",
    doi = "10.1103/PhysRevD.97.043520",
    journal = "Phys. Rev. D",
    volume = "97",
    number = "4",
    pages = "043520",
    year = "2018"
}

@article{Ferreira:2018qss,
    author = "Ferreira, Pedro G. and Hill, Christopher T. and Noller, Johannes and Ross, Graham G.",
    title = "{Inflation in a scale invariant universe}",
    eprint = "1802.06069",
    archivePrefix = "arXiv",
    primaryClass = "astro-ph.CO",
    reportNumber = "FERMILAB-PUB-18-079-T",
    doi = "10.1103/PhysRevD.97.123516",
    journal = "Phys. Rev. D",
    volume = "97",
    number = "12",
    pages = "123516",
    year = "2018"
}

@article{Harvey:1990qw,
    author = "Harvey, Jeffrey A. and Turner, Michael S.",
    title = "{Cosmological baryon and lepton number in the presence of electroweak fermion number violation}",
    reportNumber = "FERMILAB-PUB-90-049-A, EFI-90-33",
    doi = "10.1103/PhysRevD.42.3344",
    journal = "Phys. Rev. D",
    volume = "42",
    pages = "3344--3349",
    year = "1990"
}

@article{Ema:2016dny,
    author = "Ema, Yohei and Jinno, Ryusuke and Mukaida, Kyohei and Nakayama, Kazunori",
    title = "{Violent Preheating in Inflation with Nonminimal Coupling}",
    eprint = "1609.05209",
    archivePrefix = "arXiv",
    primaryClass = "hep-ph",
    doi = "10.1088/1475-7516/2017/02/045",
    journal = "JCAP",
    volume = "02",
    pages = "045",
    year = "2017"
}

@article{Sfakianakis:2018lzf,
    author = "Sfakianakis, Evangelos I. and van de Vis, Jorinde",
    title = "{Preheating after Higgs Inflation: Self-Resonance and Gauge boson production}",
    eprint = "1810.01304",
    archivePrefix = "arXiv",
    primaryClass = "hep-ph",
    reportNumber = "Nikhef-2018-044",
    doi = "10.1103/PhysRevD.99.083519",
    journal = "Phys. Rev. D",
    volume = "99",
    number = "8",
    pages = "083519",
    year = "2019"
}

@article{Gorbunov:2018llf,
    author = "Gorbunov, Dmitry and Tokareva, Anna",
    title = "{Scalaron the healer: removing the strong-coupling in the Higgs- and Higgs-dilaton inflations}",
    eprint = "1807.02392",
    archivePrefix = "arXiv",
    primaryClass = "hep-ph",
    doi = "10.1016/j.physletb.2018.11.015",
    journal = "Phys. Lett. B",
    volume = "788",
    pages = "37--41",
    year = "2019"
}

@article{Ema:2020zvg,
    author = "Ema, Yohei and Mukaida, Kyohei and van de Vis, Jorinde",
    title = "{Higgs inflation as nonlinear sigma model and scalaron as its $\sigma$-meson}",
    eprint = "2002.11739",
    archivePrefix = "arXiv",
    primaryClass = "hep-ph",
    reportNumber = "DESY 20-031, DESY-20-031",
    doi = "10.1007/JHEP11(2020)011",
    journal = "JHEP",
    volume = "11",
    pages = "011",
    year = "2020"
}

@article{Ema:2019fdd,
    author = "Ema, Yohei",
    title = "{Dynamical Emergence of Scalaron in Higgs Inflation}",
    eprint = "1907.00993",
    archivePrefix = "arXiv",
    primaryClass = "hep-ph",
    reportNumber = "DESY 19-117, DESY-19-117",
    doi = "10.1088/1475-7516/2019/09/027",
    journal = "JCAP",
    volume = "09",
    pages = "027",
    year = "2019"
}

@article{Gorbunov:2012ns,
    author = "Gorbunov, Dmitry and Tokareva, Anna",
    title = "{$R^2$-inflation with conformal SM Higgs field}",
    eprint = "1212.4466",
    archivePrefix = "arXiv",
    primaryClass = "astro-ph.CO",
    doi = "10.1088/1475-7516/2013/12/021",
    journal = "JCAP",
    volume = "12",
    pages = "021",
    year = "2013"
}

@article{Hamada:2016onh,
    author = "Hamada, Yuta and Kawai, Hikaru and Nakanishi, Yukari and Oda, Kin-ya",
    title = "{Meaning of the field dependence of the renormalization scale in Higgs inflation}",
    eprint = "1610.05885",
    archivePrefix = "arXiv",
    primaryClass = "hep-th",
    reportNumber = "OU-HET-906, KEK-TH-1926, MAD-TH-16-09",
    doi = "10.1103/PhysRevD.95.103524",
    journal = "Phys. Rev. D",
    volume = "95",
    number = "10",
    pages = "103524",
    year = "2017"
}

@article{Adelberger:2003zx,
    author = "Adelberger, E. G. and Heckel, Blayne R. and Nelson, A. E.",
    title = "{Tests of the gravitational inverse square law}",
    eprint = "hep-ph/0307284",
    archivePrefix = "arXiv",
    doi = "10.1146/annurev.nucl.53.041002.110503",
    journal = "Ann. Rev. Nucl. Part. Sci.",
    volume = "53",
    pages = "77--121",
    year = "2003"
}

@article{Li:2022ugn,
    author = "Li, Qiang and Moroi, Takeo and Nakayama, Kazunori and Yin, Wen",
    title = "{Instability of the electroweak vacuum in Starobinsky inflation}",
    eprint = "2206.05926",
    archivePrefix = "arXiv",
    primaryClass = "hep-ph",
    reportNumber = "TU-1159",
    doi = "10.1007/JHEP09(2022)102",
    journal = "JHEP",
    volume = "09",
    pages = "102",
    year = "2022"
}

@article{Kobakhidze:2013tn,
    author = "Kobakhidze, Archil and Spencer-Smith, Alexander",
    title = "{Electroweak Vacuum (In)Stability in an Inflationary Universe}",
    eprint = "1301.2846",
    archivePrefix = "arXiv",
    primaryClass = "hep-ph",
    doi = "10.1016/j.physletb.2013.04.013",
    journal = "Phys. Lett. B",
    volume = "722",
    pages = "130--134",
    year = "2013"
}

@article{Fairbairn:2014zia,
    author = "Fairbairn, Malcolm and Hogan, Robert",
    title = "{Electroweak Vacuum Stability in light of BICEP2}",
    eprint = "1403.6786",
    archivePrefix = "arXiv",
    primaryClass = "hep-ph",
    reportNumber = "KCL-PH-TH-2014-10",
    doi = "10.1103/PhysRevLett.112.201801",
    journal = "Phys. Rev. Lett.",
    volume = "112",
    pages = "201801",
    year = "2014"
}

@article{Hook:2014uia,
    author = "Hook, Anson and Kearney, John and Shakya, Bibhushan and Zurek, Kathryn M.",
    title = "{Probable or Improbable Universe? Correlating Electroweak Vacuum Instability with the Scale of Inflation}",
    eprint = "1404.5953",
    archivePrefix = "arXiv",
    primaryClass = "hep-ph",
    reportNumber = "MCTP-14-10",
    doi = "10.1007/JHEP01(2015)061",
    journal = "JHEP",
    volume = "01",
    pages = "061",
    year = "2015"
}

@article{Espinosa:2007qp,
    author = "Espinosa, J. R. and Giudice, G. F. and Riotto, A.",
    title = "{Cosmological implications of the Higgs mass measurement}",
    eprint = "0710.2484",
    archivePrefix = "arXiv",
    primaryClass = "hep-ph",
    reportNumber = "CERN-PH-TH-2007-179, IFT-UAM-CSIC-07-50",
    doi = "10.1088/1475-7516/2008/05/002",
    journal = "JCAP",
    volume = "05",
    pages = "002",
    year = "2008"
}

@article{Espinosa:2015qea,
    author = "Espinosa, Jose R. and Giudice, Gian F. and Morgante, Enrico and Riotto, Antonio and Senatore, Leonardo and Strumia, Alessandro and Tetradis, Nikolaos",
    title = "{The cosmological Higgstory of the vacuum instability}",
    eprint = "1505.04825",
    archivePrefix = "arXiv",
    primaryClass = "hep-ph",
    reportNumber = "CERN-PH-TH-2015-119",
    doi = "10.1007/JHEP09(2015)174",
    journal = "JHEP",
    volume = "09",
    pages = "174",
    year = "2015"
}

@article{Kamada:2014ufa,
    author = "Kamada, Kohei",
    title = "{Inflationary cosmology and the standard model Higgs with a small Hubble induced mass}",
    eprint = "1409.5078",
    archivePrefix = "arXiv",
    primaryClass = "hep-ph",
    doi = "10.1016/j.physletb.2015.01.024",
    journal = "Phys. Lett. B",
    volume = "742",
    pages = "126--135",
    year = "2015"
}

@article{Herranen:2014cua,
    author = "Herranen, Matti and Markkanen, Tommi and Nurmi, Sami and Rajantie, Arttu",
    title = "{Spacetime curvature and the Higgs stability during inflation}",
    eprint = "1407.3141",
    archivePrefix = "arXiv",
    primaryClass = "hep-ph",
    reportNumber = "IMPERIAL-TP-2014-AR-2",
    doi = "10.1103/PhysRevLett.113.211102",
    journal = "Phys. Rev. Lett.",
    volume = "113",
    number = "21",
    pages = "211102",
    year = "2014"
}

@article{Ema:2016kpf,
    author = "Ema, Yohei and Mukaida, Kyohei and Nakayama, Kazunori",
    title = "{Fate of Electroweak Vacuum during Preheating}",
    eprint = "1602.00483",
    archivePrefix = "arXiv",
    primaryClass = "hep-ph",
    reportNumber = "UT-16-04, IPMU-16-0012",
    doi = "10.1088/1475-7516/2016/10/043",
    journal = "JCAP",
    volume = "10",
    pages = "043",
    year = "2016"
}

@article{Figueroa:2017slm,
    author = "Figueroa, Daniel G. and Rajantie, Arttu and Torrenti, Francisco",
    title = "{Higgs field-curvature coupling and postinflationary vacuum instability}",
    eprint = "1709.00398",
    archivePrefix = "arXiv",
    primaryClass = "astro-ph.CO",
    reportNumber = "IFT-UAM-CSIC-17-078, IFT-UAM/CSIC-17-078",
    doi = "10.1103/PhysRevD.98.023532",
    journal = "Phys. Rev. D",
    volume = "98",
    number = "2",
    pages = "023532",
    year = "2018"
}

@article{Abazajian:2019eic,
    author = "Abazajian, Kevork and others",
    title = "{CMB-S4 Science Case, Reference Design, and Project Plan}",
    eprint = "1907.04473",
    archivePrefix = "arXiv",
    primaryClass = "astro-ph.IM",
    reportNumber = "FERMILAB-PUB-19-431-AE-SCD",
    month = "7",
    year = "2019"
}

@article{BICEP:2021xfz,
    author = "Ade, P. A. R. and others",
    collaboration = "BICEP, Keck",
    title = "{Improved Constraints on Primordial Gravitational Waves using Planck, WMAP, and BICEP/Keck Observations through the 2018 Observing Season}",
    eprint = "2110.00483",
    archivePrefix = "arXiv",
    primaryClass = "astro-ph.CO",
    doi = "10.1103/PhysRevLett.127.151301",
    journal = "Phys. Rev. Lett.",
    volume = "127",
    number = "15",
    pages = "151301",
    year = "2021"
}

@article{Ema:2017rqn,
    author = "Ema, Yohei",
    title = "{Higgs Scalaron Mixed Inflation}",
    eprint = "1701.07665",
    archivePrefix = "arXiv",
    primaryClass = "hep-ph",
    reportNumber = "UT-17-04",
    doi = "10.1016/j.physletb.2017.04.060",
    journal = "Phys. Lett. B",
    volume = "770",
    pages = "403--411",
    year = "2017"
}

@article{He:2018gyf,
    author = "He, Minxi and Starobinsky, Alexei A. and Yokoyama, Jun'ichi",
    title = "{Inflation in the mixed Higgs-$R^2$ model}",
    eprint = "1804.00409",
    archivePrefix = "arXiv",
    primaryClass = "astro-ph.CO",
    doi = "10.1088/1475-7516/2018/05/064",
    journal = "JCAP",
    volume = "05",
    pages = "064",
    year = "2018"
}

@article{Salvio:2014soa,
    author = "Salvio, Alberto and Strumia, Alessandro",
    title = "{Agravity}",
    eprint = "1403.4226",
    archivePrefix = "arXiv",
    primaryClass = "hep-ph",
    reportNumber = "FTUAM-14-9, IFT-UAM-CSIC-14-021",
    doi = "10.1007/JHEP06(2014)080",
    journal = "JHEP",
    volume = "06",
    pages = "080",
    year = "2014"
}

@article{Kannike:2015apa,
    author = {Kannike, Kristjan and H\"utsi, Gert and Pizza, Liberato and Racioppi, Antonio and Raidal, Martti and Salvio, Alberto and Strumia, Alessandro},
    title = "{Dynamically Induced Planck Scale and Inflation}",
    eprint = "1502.01334",
    archivePrefix = "arXiv",
    primaryClass = "astro-ph.CO",
    reportNumber = "IFT-UAM-CSIC-15-015",
    doi = "10.1007/JHEP05(2015)065",
    journal = "JHEP",
    volume = "05",
    pages = "065",
    year = "2015"
}

@article{Wetterich:1987fm,
    author = "Wetterich, C.",
    title = "{Cosmology and the Fate of Dilatation Symmetry}",
    eprint = "1711.03844",
    archivePrefix = "arXiv",
    primaryClass = "hep-th",
    reportNumber = "PRINT-87-0756, DESY-87-123",
    doi = "10.1016/0550-3213(88)90193-9",
    journal = "Nucl. Phys. B",
    volume = "302",
    pages = "668--696",
    year = "1988"
}

\end{document}